\documentclass[a4paper,11pt]{article}
\pdfoutput=1 

\usepackage{jcappub} 

\usepackage[T1]{fontenc} 
\usepackage{verbatim}
\usepackage[utf8]{inputenc}
\usepackage[american]{babel}
\usepackage{graphicx}
\usepackage{epsfig}
\usepackage{adjustbox}
\usepackage{booktabs}
\usepackage{multirow}
\usepackage{dcolumn}
\usepackage{amsmath}
\usepackage{amsfonts}
\usepackage{mathtools}
\usepackage{amssymb}
\usepackage{empheq}
\usepackage{subcaption} 
\usepackage{caption}
\usepackage{epstopdf}
\usepackage{bm}
\usepackage{mathrsfs}
\usepackage{version}
\usepackage{enumerate}
\usepackage{braket}
\usepackage{enumitem}
\usepackage{transparent}
\usepackage{pifont}
\usepackage{colortbl}
\usepackage{rotating}
\usepackage{adjustbox}
\usepackage{cancel}
\usepackage{tabularx}
\usepackage{soul}
\usepackage[table]{xcolor}
\usepackage{siunitx}
\usepackage{float} 
\usepackage[nocompress]{cite} 
\usepackage{marginnote}
\usepackage{relsize}

\usepackage{hyperref}
\hypersetup{colorlinks=true,
linkcolor=navyblue,citecolor=coralred,urlcolor=green(munsell),pdfencoding=auto}

\definecolor{navyblue}{rgb}{0.0, 0.0, 0.5}
\definecolor{bleudefrance}{rgb}{0.19, 0.55, 0.91}
\definecolor{coralred}{rgb}{1.0, 0.25, 0.25}
\definecolor{royalblue}{rgb}{0.25, 0.41, 0.88}
\definecolor{cadmiumgreen}{rgb}{0.0, 0.42, 0.24}
\definecolor{green(munsell)}{rgb}{0.0, 0.66, 0.47}
\definecolor{blue-violet}{rgb}{0.54, 0.17, 0.89}
\definecolor{darkviolet}{rgb}{0.58, 0.0, 0.83}
\definecolor{orange(colorwheel)}{rgb}{1.0, 0.5, 0.0}
\definecolor{internationalorange}{rgb}{1.0, 0.31, 0.0}

\definecolor{magenta(process)}{rgb}{1.0, 0.0, 0.56}

\definecolor{darkspringgreen}{rgb}{0.09, 0.45, 0.27}

\definecolor{royalblue(web)}{rgb}{0.25, 0.41, 0.88}

\definecolor{cadmiumorange}{rgb}{0.93, 0.53, 0.18}

\definecolor{heliotrope}{rgb}{0.87, 0.45, 1.0}

\makeatletter
\renewcommand*{\@textcolor}[3]{%
\protect\leavevmode
\begingroup
\color#1{#2}#3%
\endgroup
}
\makeatother

\newcommand{\myfloatalign}{\centering}

\DeclareCaptionLabelSeparator{colquad}{.\,\,\,}
\renewcommand\[{\left[}

\DeclarePairedDelimiter{\abs}{\lvert}{\rvert}

\makeatletter
\let\save@mathaccent\mathaccent
\newcommand*\if@single[3]{%
\setbox0\hbox{${\mathaccent"0362{#1}}^H$}%
\setbox2\hbox{${\mathaccent"0362{\kern0pt#1}}^H$}%
\ifdim\ht0=\ht2 #3\else #2\fi
}
\newcommand*\rel@kern[1]{\kern#1\dimexpr\macc@kerna}
\newcommand*\widebar[1]{\@ifnextchar^{{\wide@bar{#1}{0}}}{\wide@bar{#1}{1}}}
\newcommand*\wide@bar[2]{\if@single{#1}{\wide@bar@{#1}{#2}{1}}{\wide@bar@{#1}{#2}{2}}}
\newcommand*\wide@bar@[3]{%
\begingroup
\def\mathaccent##1##2{%
\let\mathaccent\save@mathaccent
\if#32 \let\macc@nucleus\first@char \fi
\setbox\z@\hbox{$\macc@style{\macc@nucleus}_{}$}%
\setbox\tw@\hbox{$\macc@style{\macc@nucleus}{}_{}$}%
\dimen@\wd\tw@
\advance\dimen@-\wd\z@
\divide\dimen@ 3
\@tempdima\wd\tw@
\advance\@tempdima-\scriptspace
\divide\@tempdima 10
\advance\dimen@-\@tempdima
\ifdim\dimen@>\z@ \dimen@0pt\fi
\rel@kern{0.6}\kern-\dimen@
\if#31
\overline{\rel@kern{-0.6}\kern\dimen@\macc@nucleus\rel@kern{0.4}\kern\dimen@}%
\advance\dimen@0.4\dimexpr\macc@kerna
\let\final@kern#2%
\ifdim\dimen@<\z@ \let\final@kern1\fi
\if\final@kern1 \kern-\dimen@\fi
\else
\overline{\rel@kern{-0.6}\kern\dimen@#1}%
\fi
}%
\macc@depth\@ne
\let\math@bgroup\@empty \let\math@egroup\macc@set@skewchar
\mathsurround\z@ \frozen@everymath{\mathgroup\macc@group\relax}%
\macc@set@skewchar\relax
\let\mathaccentV\macc@nested@a
\if#31
\macc@nested@a\relax111{#1}%
\else
\def\gobble@till@marker##1\endmarker{}%
\futurelet\first@char\gobble@till@marker#1\endmarker
\ifcat\noexpand\first@char A\else
\def\first@char{}%
\fi
\macc@nested@a\relax111{\first@char}%
\fi
\endgroup
}
\makeatother

\newcommand\ee{\end{equation}}
\newcommand\be{\begin{equation}}
\newcommand\eea{\end{eqnarray}}
\newcommand\bea{\begin{eqnarray}}
\newcommand{\bsp}{\begin{split}}
\newcommand{\esp}{\end{split}}
\newcommand{\bit}{\begin{itemize}[leftmargin=*]}
\newcommand{\eit}{\end{itemize}}
\newcommand{\ben}{\begin{enumerate}[leftmargin=*]}
\newcommand{\een}{\end{enumerate}}

\newcommand{\ie}{
i.e.
}
\newcommand{\eg}{
e.g.
}

\newcommand{\cmark}{\ding{51}}
\newcommand{\xmark}{\ding{55}}

\newcommand{\kmpc}[1]{\SI{#1}{\mathrm{Mpc}^{-1}}}

\newcommand{\kmpch}[1]{\num{#1}{\,h\,\mathrm{Mpc}^{-1}}}
\newcommand{\lmpch}[1]{\num{#1}{\,h^{-1}\,\mathrm{Mpc}}}
\newcommand{\solar}[1]{\num{#1}{\,h^{-1}M_\odot}}

\newcommand\eq[1]{Eq.~\eqref{eq:#1}}

\newcommand{\eqsII}[2]{Eqs.~\eqref{eq:#1}, \eqref{eq:#2}}
\newcommand{\eqsIII}[3]{Eqs.~\eqref{eq:#1}, \eqref{eq:#2}, \eqref{eq:#3}}

\newcommand{\eqsV}[5]{Eqs.~\eqref{eq:#1}, \eqref{eq:#2}, \eqref{eq:#3}, \eqref{eq:#4}, \eqref{eq:#5}}

\newcommand\fig[1]{Fig.~\ref{fig:#1}}
\newcommand\tab[1]{Tab.~\ref{tab:#1}}
\newcommand\figs[2]{Figs.~\ref{fig:#1},~\ref{fig:#2}}
\newcommand\tabs[2]{Tabs.~\ref{tab:#1},~\ref{tab:#2}}

\newcommand{\dif}{\mathrm{d}}

\renewcommand{\vec}{\bm} 
\newcommand\vers[1]{\hat{\vec{#1}}}

\newcommand{\HI}{{\text{HI}}}
\newcommand{\DIV}{~{\text{--}}~}
\newcommand{\fHI}{f_{\epsilon_{\HI}}}

\def\comment#1{}

\title{A new scale in the bias expansion} 

\author[a]{Giovanni Cabass,}
\author[a]{Fabian Schmidt}

\affiliation[a]{Max-Planck-Institut f\"{u}r Astrophysik, 
Karl-Schwarzschild-Str. 1, 85741 Garching, Germany}

\emailAdd{gcabass@mpa-garching.mpg.de}
\emailAdd{fabians@mpa-garching.mpg.de}

\abstract{\noindent The fact that the spatial nonlocality of galaxy formation is controlled by some short length scale 
like the Lagrangian radius is the cornerstone of the bias expansion for large-scale-structure tracers. 
However, the first sources of ionizing radiation between $z\approx 15$ and $z\approx 6$ 
are expected to have significant effects on the formation of galaxies we 
observe at lower redshift, at least on low-mass galaxies. These 
radiative-transfer effects introduce a new scale in the clustering of galaxies, 
i.e.~the finite distance which ionizing radiation travels until it reaches a given galaxy. 
This mean free path can be very large, of order $100\,h^{-1}\,\mathrm{Mpc}$. 
Consequently, higher-derivative terms in the bias expansion could turn out to be non-negligible even on these scales: 
treating them perturbatively would lead to a massive loss in predictivity and, for example, 
could spoil the determination of the BAO feature or constraints on the neutrino mass. Here, 
we investigate under what assumptions an explicit non-perturbative model of radiative-transfer effects 
can maintain the robustness of large-scale galaxy clustering as a cosmological probe.} 

\begin{document}
\maketitle
\flushbottom


\section{Introduction}
\label{sec:introduction}

\noindent A crucial step in connecting cosmological scenarios with large-scale-structure (LSS) 
observations is the bias expansion for LSS tracers (see \cite{Desjacques:2016bnm} for a review). 
Like all effective field theories, the bias expansion is firmly rooted in the idea of locality. The 
simplest example is that of dark matter halos. Since their dynamics is governed by gravitational interactions only, 
we know that at any given point in spacetime the halo overdensity $\delta_h$ will be a functional of all possible gravitational observables 
that one can build from the Newtonian potential. Let us focus, for example, on the 
total matter overdensity $\delta$. At leading order in perturbation theory, we can write the most general 
functional dependence of the halo overdensity on $\delta$ as the integral 
\begin{equation}
\label{eq:locality-A}
\delta_h(\eta,\vec{x})=\int\dif\eta'\dif^3y\,F_h(\eta,\eta',\abs{\vec{y}})\,\delta(\eta',\vec{x}+\vec{y})\,\,,
\end{equation}
where the kernel $F_h$ can only depend on $\abs{\vec{y}}$ because of statistical isotropy, 
and cannot depend on $\vec{x}$ because of statistical homogeneity. Even 
if we do not know the exact shape of the kernel, we know that it is supported only for $\abs{\vec{y}}\lesssim R(M_h)$, 
where $R(M_h)$ is the Lagrangian radius of the halo. This 
is because the matter within a given halo originates from a region of size $R(M_h)$ in Lagrangian space. We 
can then expand $\delta(\eta',\vec{x}+\vec{y})$ in a Taylor series inside the integral, and obtain 
\begin{equation}
\label{eq:locality-B}
\delta_h(\eta,\vec{x}) = \int\dif\eta'\dif^3y\,F_h(\eta,\eta',\abs{\vec{y}})\,\delta(\eta',\vec{x}) 
+ \frac{1}{6}\int\dif\eta'\dif^3y\,F_h(\eta,\eta',\abs{\vec{y}})\,\abs{\vec{y}}^2\,\nabla^2\delta(\eta',\vec{x}) + \dots\,\,.
\end{equation}
Finally, using the fact that, at linear order in perturbations, $\delta$ evolves in a scale-independent way, 
we can easily carry out the time integration to arrive at the expansion 
\begin{equation}
\label{eq:locality-C}
\delta_h(\eta,\vec{x}) = b_1(\eta)\delta(\eta,\vec{x})+b_{\nabla^2\delta}(\eta)\nabla^2\delta(\eta,\vec{x})+\dots\,\,,
\end{equation}
where the bias coefficients $b_1,b_{\nabla^2\delta},\dots$, are related to the moments of the kernel $F_h$ \cite{McDonald:2009dh}. Since 
the nonlocality scale of the kernel is $R(M_h)$, we see that $b_{\nabla^{2n}\delta}\sim R^{2n}(M_h)$. 

So far, so good: everything seems to be controlled by the single scale $R(M_h)$. 
Moreover, this scale is typically of the same order of magnitude as the scale at which the matter density field becomes nonlinear, 
so that it does not strongly restrict the validity of the perturbative bias expansion. 
When we move from halos to galaxies and consider their overdensity $\delta_g$, 
we can still write it as in \eq{locality-C}. However, now we have to ask what is the nonlocality scale of the higher-derivative terms for galaxies. If 
the properties of galaxies in a given sample are completely determined by those of their host halos, then this scale is still $R(M_h)$. 
However, the real universe is not so simple. For example, baryons are also present. What is their impact on 
\eq{locality-C}? Pressure forces contribute to the right-hand side of the Euler equation for baryonic matter 
through a pressure gradient $\vec{\nabla}\delta p_{b} = c^2_{\rm s}\widebar{\rho}_{b}\vec{\nabla}\delta_{b}
\approx c^2_{\rm s}\widebar{\rho}_{b}\vec{\nabla}\delta$ (approximating $\delta_{b}\approx\delta$ on large scales), 
and then give rise to a baryon-dark matter relative velocity $\vec{v}_{bc}$. Since 
this relative velocity is a local observable, it can enter in the bias expansion. 
At leading order in perturbations we can then add the term $\vec{\nabla}\cdot\vec{v}_{bc}\sim\nabla^2\delta$ to \eq{locality-C}. 
Thus, we see that these baryonic effects are also captured by higher-derivative terms. 
However, the length scale suppressing them is not $R(M_h)$, but the Jeans length $\lambda_{\rm J} = c_{\rm s}(G_{\rm N}\rho)^{-1/2}$, 
which depends on the average density of gas in the halo and its speed of sound $c^2_{\rm s}\sim T/m$ (for gas particles of mass $m$). 
Fortunately, this length scale is again quite small, and less than $R(M_h)$ except for very low-mass halos. 

Another source of nonlocality is radiative transfer (RT). Indeed, 
ionizing radiation can affect star-forming galaxies directly: for example, Refs.~\cite{Efstathiou:1992zz,Barkana:1999apa} showed that 
it can reduce the cooling rate of the gas accreting onto the parent halo (effectively evaporating it from the halo), 
slowing down the star-formation rate and leading to a suppression of the stellar-mass-to-halo-mass ratio in low-mass galaxies. In 
this case the nonlocality scale is some ``effective'' mean free path (m.f.p.) $\lambda_{\rm eff}$ of ionizing radiation, 
that in the following will always be understood as the comoving one (we use the term ``effective'' since we will see that what matters 
for the bias expansion is some average of the mean free path over the photon energy). 

Both these effects are expected to become relevant during reionization, 
when the progenitors of galaxies observed at lower redshifts were actively forming \cite{Schmidt:2017lqe}. 
It is during this epoch (between redshifts $z\approx 15$ and $z\approx 6$) that newly formed radiating objects like metal-poor stars, supernova explosions, 
accreting black holes, X-ray binaries, mini-quasars, dwarf galaxies, etc. injected photons into the intergalactic medium, 
and the universe reverted from neutral to ionized. The gas temperature jumped from a few $\rm K$ to several thousands, 
leading to a Jeans length $\lambda_{\rm J}\approx\lmpch{0.08}$ around the end of reionization \cite{Schmidt:2017lqe}, 
while the m.f.p.~can be of order $\lmpch{50}$ at redshift $z\sim 5\DIV6$ \cite{Worseck:2014fya,Becker:2014oga}. 

Once we have the expansion of \eq{locality-C}, and similar expansions at higher order in perturbation theory, 
we can predict the statistics of galaxies in terms of those of the underlying matter field at a given order in perturbation theory. 
In the best case, this prediction holds up to the nonlinear scale, $k_{\rm NL}\approx\kmpch{0.3}$ at $z=0$, 
that controls the perturbation theory for the dark matter distribution. However, 
it is clear that we must be able to truncate the series in \eq{locality-C} at some order if we want to be predictive. This is easy to see in Fourier space. Let 
us consider for example the higher-derivative terms coming from RT effects. 
In \eq{locality-C} we obtain a series of terms that scale as $(k^2\lambda_{\rm eff}^2)^n$ times a coefficient that, 
naively, we expect to be of order $1$: then, the derivative expansion will break down at the very latest at $k\sim 1/\lambda_{\rm eff}$, 
since at that scale infinitely many terms will become equally relevant. 
In this paper we will focus on these terms, answering the following questions. \emph{Is it possible, within the effective field theory framework, 
to resum them in a way that allows us to predict the galaxy statistics also at momenta $k\gtrsim 1/\lambda_{\rm eff}$ (but still $k \ll k_{\rm NL}(z)$)? Can 
we do this with only a finite number of new bias coefficients? Which assumptions are necessary to achieve this?} 

These questions are important because $\lambda_{\rm eff}$ can be much larger than $\lambda_{\rm J}$: while 
there is no problem in treating the higher-derivative terms from pressure forces perturbatively (unless their bias coefficients are unnaturally large), 
doing so with those from RT can lead to a sizeable loss in predictivity. 
How does this happen? When we marginalize over the free coefficients of the bias expansion, we need to take into account their prior. 
The coefficients $b_{\nabla^{2n}\delta}$ of the higher-derivative terms are dimensionful: 
however, once we identify the longest non-locality scale $\lambda$ (be it $R(M_h)$, or $\lambda_{\rm eff}$, etc.) that we think affects the formation of the tracers under 
consideration, it is reasonable to assume that after we factor it out the dimensionless coefficients are of order $1$, as we discussed below \eq{locality-C}.\footnote{Therefore, 
we would take the priors on the coefficients $b_{\nabla^{2n}\delta}$ to be $[{-{\cal O}}(1),{\cal O}(1)]\times\lambda^{2n}$, where $\lambda^{2n}$ is fixed.} 
Consequently, if we want to keep a finite number of higher-derivative terms (and then a finite number of coefficients to marginalize over), 
we are forced to stop at a $k_{\rm max}$ which is not larger than $\sim 1/\lambda$. The longer $\lambda$, then, 
the smaller is the number of modes that we are using, and this would lead to a degradation of the constraints on cosmological parameters 
(for more details we refer, for example, to the discussions in Sections 3 and 4 (especially Section 4.2) of \cite{Gleyzes:2016tdh}). 
More precisely, these higher-derivative terms will strongly modify the shape of the galaxy two-point correlation function around BAO scales, 
damping the amplitude of the BAO feature and possibly affecting its measured position 
(see \eg~\cite{Pritchard:2006ng,Coles:2007be,Pontzen:2014ena,Gontcho:2014nsa} for a discussion). 
Finally, they will also have a negative impact on the constraints on parameters like the neutrino 
mass and the amplitude of equilateral primordial non-Gaussianity: indeed, 
the scale-dependent effects induced by neutrinos and equilateral non-Gaussianity are controlled by the free-streaming scale $k_{\rm fs}$ 
and the equality scale $k_{\rm eq}$ respectively, and both these scales could be close to $1/\lambda_{\rm eff}$. 

Before proceeding, we emphasize that these RT effects also affect line emission from diffuse gas, 
like the Lyman-$\alpha$ forest and {$\rm 21cm$} intensity mapping (see \eg~\cite{Pontzen:2014ena,Gontcho:2014nsa}). 
Actually, we expect that the response of galaxies to ionizing radiation is very suppressed if 
they reside in halos with mass much larger than the Jeans mass at reionization \cite{Schmidt:2017lqe}, 
while the line emission depends strongly on the ionization state of the medium and then on the ambient radiation field. 
This distinction is not important for this paper: since we follow an effective field theory approach, all our conclusions will 
apply to any physical tracer of the matter field (the differences between various tracers being encoded in their respective bias coefficients). 

Our paper is subdivided in four main sections. We formulate in more detail the above questions in Section \ref{sec:questions}, 
and we answer them in Sections \ref{sec:radiative_bias} and \ref{sec:full_RT}. We draw our conclusions in Section \ref{sec:conclusions}. 
Technical details on Sections \ref{sec:radiative_bias} and \ref{sec:full_RT} are collected in Appendix \ref{app:radiative_transfer}. 

\paragraph{Notation and conventions} 
We largely follow the notation of \cite{Desjacques:2016bnm}. Some differences are that we denote conformal time by $\eta$ ($\tau$ is reserved for the optical depth), 
and denote the leading stochastic term in the bias expansion for galaxies by $\epsilon_{g}$ (instead of simply $\epsilon$), since we will be dealing with multiple tracers. 
For simplicity we omit the dependence of the fluid trajectory $\vec{x}_{\rm fl}$ on the Lagrangian coordinate. 
The metric signature is $(-,+,+,+)$. We work in natural units $c=\hbar=\varepsilon_0=1$, where $\varepsilon_0$ is the permittivity of free space. Consequently, 
we freely exchange energy with (angular) frequency and time with length in our discussion of radiative transfer.

\section{Setting up the analysis}
\label{sec:questions}

\noindent As one can easily imagine, we can capture the effect of RT on galaxy formation 
by allowing their number density to be a functional of the incoming flux of ionizing radiation \cite{Schmidt:2017lqe}. In 
principle, galaxy formation is sensitive to this flux in a region of finite size, corresponding to the extent of the gas cloud around the parent halo, 
which is of order of the Jeans length of the gas. This nonlocality can then be dealt with via a derivative expansion like that of \eq{locality-A}, 
and at leading order it is enough to consider the flux evaluated along the trajectory $\vec{x}_{\rm fl}$ of a Lagrangian patch enclosing the tracer: 
higher-order corrections involve derivatives of the incoming flux along this fluid trajectory. 
The additional scales of nonlocality in the problem are the following: 
\begin{description}[leftmargin=*]
\item[Mean free path of radiation] As we discussed in the introduction, one is $\lambda_{\rm eff}$. Radiation 
can travel long distances before reaching the galaxy 
(while in contrast matter and biased tracers typically move $\lesssim \pi/k_{\rm NL}(z=0) \sim \lmpch{10}$ over the entire history of the universe). Therefore, 
since the emissivity of the sources of photoionizing radiation is also biased with respect to matter, 
we can conclude that the galaxies are now sensitive to the distribution of matter within their whole past light cone. 
However, radiation is also being absorbed by the intergalactic medium, and then 
sources farther than $\lambda_{\rm eff}$ from a given galaxy are not able to influence it. 
This is why we can assume that higher-derivative operators from RT effects are suppressed by $1/\lambda_{\rm eff}$, 
\ie~$b_{\nabla^{2n}\delta}\sim f_{(2n)}\lambda^{2n}_{\rm eff}$. With 
this assumption on the scaling of higher-derivative bias coefficients, Ref.~\cite{Schmidt:2017lqe} obtained a constraint on the bias coefficient of the operator $\nabla^2\delta$ 
for the galaxies of the BOSS DR12 galaxy sample \cite{Beutler:2016ixs,Beutler:2016arn}: $\abs{f_{(2)}}\lesssim\num{0.002}\,(\lambda_{\rm eff}/\lmpch{50})^{-2}$ 
(the smallness of $f_{(2)}$ is consistent with the fact that the galaxies of the BOSS sample reside in massive halos, $M_{h}\approx\solar{d13}$). 
\item[Response history to incoming flux] It is possible that galaxy formation does not respond instantaneously to the incoming radiation, 
but keeps some ``memory'' of it. 
To see the consequences of this, imagine for a moment that all ionizing radiation is emitted 
in an interval $\Delta\eta$ around some redshift $z_\ast$, with $\Delta\eta\ll1/{\cal H}(z_\ast)$. 
If the response of galaxies to the incoming flux of ionizing radiation was instantaneous, the galaxy number density at any event along the fluid worldline 
would only depend on the distribution of sources at the intersection of the past light cone of that event with the hypersurface $z=z_\ast$ 
(see \fig{geometry_2D} below). In general however, it also knows about sources at $z_\ast$ that are inside the past light cone, 
closer to the spatial position of the galaxies under consideration by an amount controlled by the response time. 
Clearly this effect goes in the opposite direction with respect to that of a finite mean free path of radiation, and must also be accounted for. 
\end{description}

Before discussing how to account for both these scales in our effective field theory approach, 
let us also briefly think about the alternative to this. Obviously, 
one could attempt a direct modeling of what would be called ``UV physics'' in the language of particle physics. 
This means both a model of the response of galaxies to the ionizing photons, 
but also of the spatial distribution of the radiation sources within the light cone of the observed galaxies. 
However, it is clear that we cannot directly test our models for the latter against observations, 
since the past light cone of the observed galaxies lies \emph{within} (and not \emph{on}) 
our past light cone: while in principle we could imagine to reconstruct the response of galaxies with observations at different redshifts 
(since by statistical homogeneity the response cannot depend on the spatial position of the galaxies), 
we cannot receive the light from all the sources inside it. This is easy to see in an FLRW universe 
(see for example \figs{geometry_1D}{geometry_2D} in the next section) 
or any conformally flat spacetime, but holds non-perturbatively.\footnote{In general, 
consider an event $P$ and an event $Q$ in the causal past of $P$. 
Then, any event $R$ in the causal past of $Q$ belongs to the causal past of $P$, since we can always find a timelike or lightlike curve connecting $P$ with $R$. 
However, it is clear that there are many events in the causal past of $Q$ that are not on the light cone 
of $P$, like for example any event that is reached by a past-directed timelike curve from $Q$.} 

Instead of pursuing this route, we attempt to remain as general as possible and only 
assume that the sources of ionizing radiation can be described by a bias expansion whose nonlocality scale 
is much shorter than the m.f.p.~of ionizing radiation. Further, since in this paper we are only interested in treating the higher-derivative 
terms controlled by $\lambda_{\rm eff}$, we will drop all higher-derivative terms that are suppressed by these additional short scales 
and stop at first order in perturbations. The extension to higher orders in perturbations is well understood when spatial nonlocality 
is expanded following \eq{locality-B} (see Sections 2 and 4 of \cite{Desjacques:2016bnm}). Moreover, 
we expect RT effects to typically be of relatively small amplitude, so that a linear treatment of their contribution is appropriate.

\section{Radiative-transfer effects and the bias expansion}
\label{sec:radiative_bias}

\noindent We start by presenting the equations of radiative transfer, and discuss in more detail the properties of the emission and absorption coefficients 
(Section \ref{ssec:emission_absorption}). 

Then, in order to set the stage, in Section \ref{ssec:single_flash} we consider a ``flash'' of radiation emitted 
around $\eta=\eta_\ast$, \ie~emitted within an interval $\Delta\eta$ much shorter than a Hubble time, and focus only 
on inhomogeneities in the radiation field on this hypersurface (\ie~we neglect the effect of inhomogeneities along the photon geodesics leading to the galaxies). 
We show that, in this idealized scenario, we can capture the RT effects \emph{without
expanding in derivatives} by adding one new function of $k^2$ in the bias expansion: this 
function encodes the response of the galaxies to the incoming flux of ionizing radiation along the past fluid worldline. 
We also show that, even if we do not know this response exactly, it is possible to predict the shape of this function 
at every order in an expansion in powers of the radiation m.f.p.~divided by the Hubble radius. That is, as long as the m.f.p.~is significantly smaller than the horizon, 
we can successfully resum the power series in $k^2 \lambda_{\rm eff}^2$ into a computable function. 
Our ignorance of the small-scale physics is then parameterized by the overall coefficients of this expansion, which are generic functions of time. 

This statement is not trivial. It is well known that, in the standard bias expansion, all the Green's functions that describe the nonlocality in time of galaxy formation can, 
at each order in perturbation theory, be rewritten in terms of a finite number of time-dependent (but not scale-dependent) bias coefficients. At linear order, this 
is shown explicitly in Section \ref{sec:introduction} (compare \eq{locality-B} with \eq{locality-C}). In our case, 
instead, any time dependence along the past fluid trajectory turns into a dependence on the spatial distribution of sources at $\eta=\eta_\ast$ 
(since the integrated incoming flux receives contributions from everything along the past light cone of the galaxies), 
and this leads to a modification of all the higher-derivative terms in the bias expansion. 

However, as long as the m.f.p.~is much shorter than the Hubble radius, photons are actually arriving from a small comoving volume 
of size $\sim\lambda_{\rm eff}^3$ around the past fluid trajectory. This is what ultimately allows a resummation 
of all these higher-derivative terms into specific functions of $k^2$, each proportional to an increasing power of ${\cal H}\lambda_{\rm eff}$. Crucially, 
this resummation allows us to describe galaxy clustering even for $k\gg 1/\lambda_{\rm eff}$. We will later see whether this also holds beyond 
the instantaneous flash and homogeneous medium approximations. 

We compute the impact of the resummed RT effects on the galaxy power spectrum and 
possible degeneracies with the neutrino mass in Section \ref{ssec:galaxy_statistics}. 
In Sections \ref{ssec:q_tracers} and \ref{ssec:power_spectra} we discuss briefly how to treat tracers 
that are immune to the very nonlocal effects discussed above (we label these tracers by ``$q$'' in the following). 
This provides an illustration for how multiple LSS tracers might help in understanding the RT contributions.

\subsection{Radiative-transfer equation} 
\label{ssec:emission_absorption}

\noindent Let us first write down the equation of radiative transfer. 
We define the phase-space density of emitted radiation as $\mathcal{N}(x^\mu,P^\nu)$, where $P^\mu$ is the photon four-momentum. Calling $U^\mu$ 
the four-velocity of the observer that follows the fluid worldline $x^\mu_{\rm fl} = (\eta,\vec{x}_{\rm fl}(\eta))$, 
we can decompose $P^\mu$ as $E(U^\mu+l^\mu)$, where $E$ and $l^\mu$ are, respectively, 
the photon energy and its direction as measured by the observer defined by $U^\mu$. Consequently, 
we can define the specific intensity of emitted radiation for the observer $U^\mu$ as ${\cal I} = E^3 \mathcal{N}$. 
Its evolution is dictated by the Boltzmann equation \mbox{(see \eg~\cite{Rybicki:2004hfl})} 
\begin{equation}
\label{eq:boltzmann-1}
\frac{{\rm D}{\cal I}}{\dif\lambda} = \frac{3{\cal I}}{E}\frac{{\rm D} E}{\dif\lambda} + {\cal I}\sigma_{\rm ab} j^\mu_{\rm ab} P_\mu 
- \frac{\rho_{\rm em}\varepsilon_{\rm em}{U^\mu_{\rm em}}P_\mu}{4\pi}\,\,, 
\end{equation}
where $\lambda$ is an affine parameter along the photon geodesics, normalized such that $P^\mu = {\dif x^\mu}/{\dif\lambda}$, 
and the term $3{\cal I}\times({\rm D}\log E/{\rm d}\lambda)$ encodes, \eg, the dilution due to the expansion of the universe. For simplicity of 
notation we have suppressed the argument of $\cal I$: in general, like $\cal N$, it is also a function of the spacetime position $x^\mu$ and the photon four-momentum $P^\mu$. 

Let us discuss in more detail the second and the third term on the right-hand side of \eq{boltzmann-1}, which are directly related to the so-called emission and absorption coefficients. 
For simplicity, here we have considered a single family of emitter and absorbers, since the generalization to an arbitrary number is straightforward: 
\begin{itemize}
\item the number current of absorbers (\eg~neutral hydrogen) is given by $j^\mu_{\rm ab} = n_{\rm ab} U^\mu_{\rm ab}$, 
where $U^\mu_{\rm ab}$ is their four-velocity and $n_{\rm ab}$ their number density; 
\item the absorption cross section is called $\sigma_{\rm ab}$. 
In the case of absorption by neutral hydrogen with emission of an electron in the continuum (photoionization), this would be the bound-free cross section $\sigma_{\rm bf}$; 
\item $U^\mu_{\rm em}$ is the four-velocity of the emitting medium and $\rho_{\rm em}$ its mass density. 
For emitters of fixed mass $m_{\rm em}$ and number current $j^\mu_{\rm em}$ we have $\rho_{\rm em} U^\mu_{\rm em} = m_{\rm em} j^\mu_{\rm em}$, 
\item the dimensionless function $\varepsilon_{\rm em}(x^\mu,P^\nu)$ is the emissivity, 
defined as the energy emitted by the source medium per unit frequency per unit time per unit mass \cite{Rybicki:2004hfl}. 
\end{itemize}
With this, the emission and absorption coefficients, 
defined as the energy emitted per unit time per unit solid angle per unit volume and the cross-sectional area presented by absorbers per unit volume, 
are $(\rho_{\rm em}\varepsilon_{\rm em})/(4\pi)$ and $\sigma_{\rm ab}n_{\rm ab}$, respectively.

What about the absorption cross section and the emissivity? There is no preferred direction towards which the constituents 
of the source medium can align (at first order in perturbations): therefore the emissivity does not depend on the photon direction $l^\mu$. Moreover, 
here we are also considering the total absorption cross section, so that $\sigma_{\rm ab}$ depends only on the photon energy.\footnote{Consider 
for example the photoionization of hydrogen in the $1s$ state. If 
the emitted electron is nonrelativistic, the angular dependence of the differential cross section is given by $\abs{\vers{k}_e\cdot\vers{\epsilon}}^2$, 
where $\vec{k}_e$ is the momentum of the outgoing electron and $\vers{\epsilon}$ is the polarization of the incoming photon. 
However, in this case we are not interested in 
the direction of the outgoing electron, so we integrate over it (spin is conserved for nonrelativistic electrons, so tracing over it is trivial). 
Further, we average over $\vers{\epsilon}$ since we consider unpolarized radiation.} 
We emphasize that these assumptions are common to both analytic studies of reionization \cite{Mao:2014rja} 
and radiative-transfer simulations (see \eg~the discussion in \cite{McQuinn:2018zwa}). 
Finally, $\varepsilon_{\rm em}$ does not depend on the fluid trajectory $\vec{x}_{\rm fl}(\eta)$ but only on time, since 
the emission of radiation is localized at the position of the sources. For some examples of specific source models we refer to \cite{Zhang:2006kr,DAloisio:2013mgn,Mao:2014rja}. 

Before proceeding, we also emphasize that in \eq{boltzmann-1} we have neglected scattering. 
We discuss this issue in more detail in Section \ref{sec:full_RT}. While \eq{boltzmann-1} holds in general spacetimes, 
for the purposes of this paper it is sufficient to consider an FLRW universe, and neglect metric perturbations. 
More precisely, lensing will be relevant only at second order in perturbations (since a nontrivial lensing effect only arises if there are anisotropies in the radiation field), 
and gravitational redshift effects (Sachs-Wolfe and Integrated Sachs-Wolfe) can be neglected on sufficiently sub-horizon scales. 

Let us now take advantage of the fact that, if we neglect higher-derivative terms suppressed by the halo Lagrangian radius, 
all of emitters, absorbers, and receivers are comoving with the matter fluid, \ie~there is no velocity bias 
\cite{Senatore:2014eva,Mirbabayi:2014zca,Desjacques:2016bnm}. 
Then, we have that $j^\mu_{\rm ab} = n_{\rm ab} U^\mu$ and $U^\mu_{\rm em} = U^\mu$. 
This tells us that $j^\mu_{\rm ab} P_\mu = {-n_{\rm ab}}E$, and ${-P_\mu U_{\rm em}^\mu} = E$. 
Therefore \eq{boltzmann-1} becomes 
\begin{equation}
\label{eq:boltzmann-2}
E(U^\mu+l^\mu)\nabla_\mu{\cal I} + \frac{\partial{\cal I}}{\partial E}\frac{{\rm D}E}{{\rm d}\lambda} 
= \frac{3{\cal I}}{E}\frac{{\rm D} E}{\dif\lambda} - {\cal I}\sigma_{\rm ab} n_{\rm ab} E + \frac{\rho_{\rm em}\varepsilon_{\rm em}E}{4\pi}\,\,,
\end{equation}
where we have expanded ${\rm D}/{\rm d}\lambda$ using the fact that ${\cal I}$ does not depend explicitly on $l^\mu$, 
given our assumptions on $\sigma_{\rm ab}$ and $\varepsilon_{\rm em}$. 
Then, for an FLRW metric in comoving coordinates 
we have $l^\mu=(0,-\vers{n}/a)$, where $\vers{n}$ is opposite to the photon direction: 
$\vers{n}$ remains constant because there is no lensing, and ${\rm D}E/{\rm d}\lambda = -HE^2$. Moreover, 
since we stop at first order in perturbations in the bias expansion, it is not necessary to consider displacements \cite{Senatore:2014eva,Mirbabayi:2014zca,Desjacques:2016bnm}: 
that is, the fluid worldline is just given by $x^\mu_{\rm fl} = (\eta,\vec{x}_{\rm fl}(\eta)) = (\eta,\vec{x})$ and then $U^\mu=\delta^\mu_0/a$. 
\eq{boltzmann-2} then becomes
\begin{equation}
\label{eq:boltzmann-3}
\frac{\partial{\cal I}}{\partial\eta} - \vers{n}\cdot\vec{\nabla}{\cal I} - {\cal H}E\frac{\partial{\cal I}}{\partial E} = 
{-3}{\cal H}{\cal I} - {\cal I}\sigma_{\rm ab} n_{\rm ab}a + \frac{\rho_{\rm em}\varepsilon_{\rm em}a}{4\pi}\,\,, 
\end{equation}
where now and in the following we write ${\cal I}$ in terms of the arguments ${\cal I}(\eta,\vec{x},E,\vers{n})$.\footnote{Given our assumptions of isotropic emissivity and no scattering the dependence on $\vers{n}$ 
comes only from the photon free-streaming after emission: we refer to \eq{boltzmann_solution} for details.} 

It is straightforward to solve \eq{boltzmann-3} by an integral along the line of sight. This is done in Appendix \ref{app:solution_of_boltzmann_equation}. 
Before discussing the most general solution (which will be done in Section \ref{sec:full_RT}), let us focus on the case 
where the emissivity is non-vanishing only for a short interval of time around $\eta_\ast$, 
and there are no inhomogeneities in the number density of absorbers ($n_{\rm ab}(\eta,\vec{x}) = \widebar{n}_{\rm ab}(\eta)$). 
Clearly, the time evolution of the emissivity and the fluctuations in $n_{\rm ab}$ must also be considered. 
However, these assumptions allow us to introduce in a simple way the response of galaxies to the ionizing radiation. 
Moreover, as discussed above, we will show that 
it is in this scenario that we are able to predict how the correction to the galaxy bias from radiative transfer depends on $k$.

\subsection{Instantaneous flash and homogeneous optical depth}
\label{ssec:single_flash}

\begin{figure}
\centering
\includegraphics[width=0.85
\columnwidth]{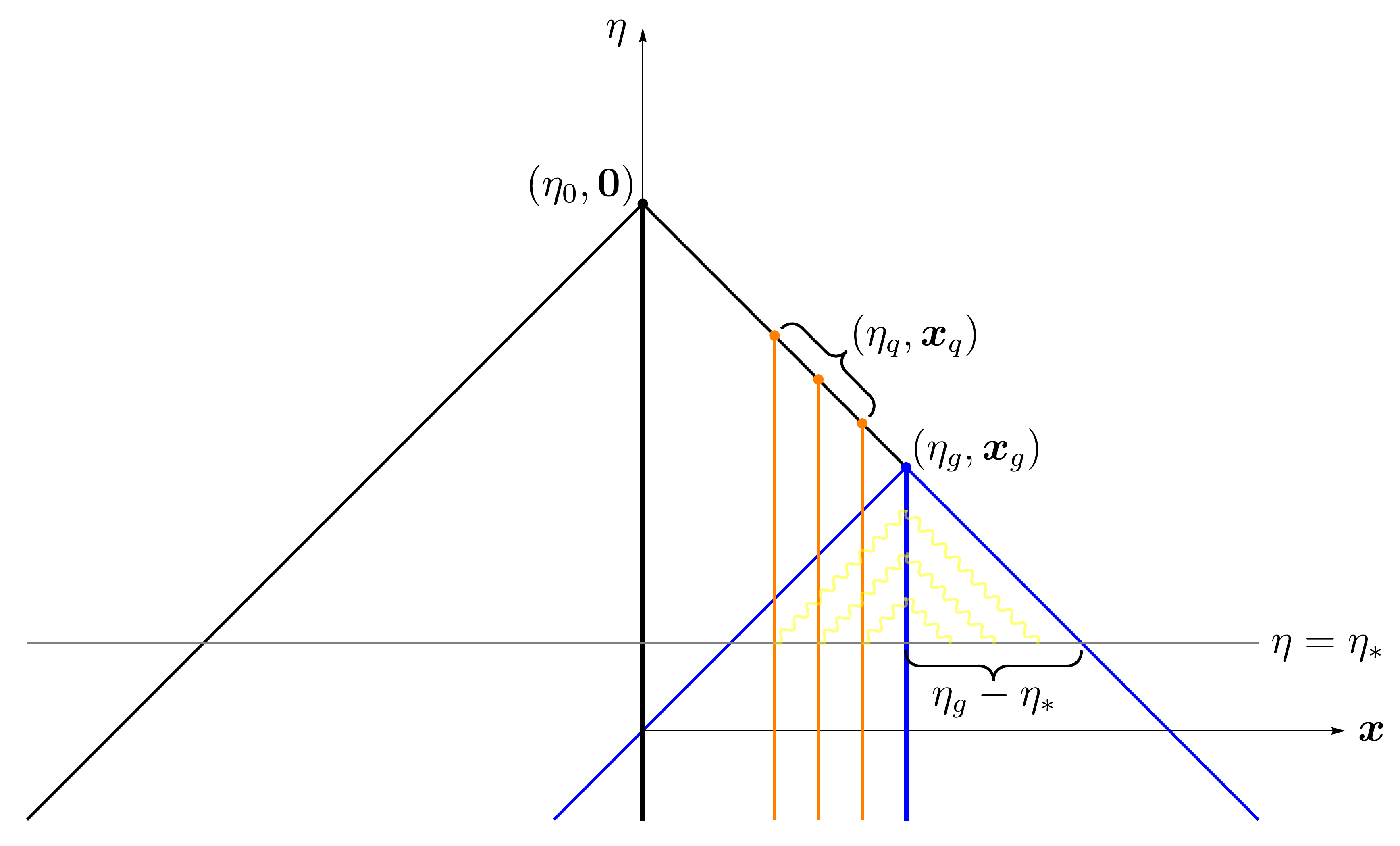} 
\caption{Spacetime diagram for a single emission time $\eta_\ast$. The observer is at $(\eta_0,\vec{0})$, 
while the observed galaxies are at $(\eta_{g},\vec{x}_{g})$. The tracers that we assume to be locally biased 
with respect to the radiation field are observed at $(\eta_{q},\vec{x}_{q})$.} 
\label{fig:geometry_1D}
\end{figure}

\noindent The spacetime diagram that summarizes our setup is shown in \figs{geometry_1D}{geometry_2D}. 
Radiation is emitted in a single flash around $\eta=\eta_\ast$, and can affect the galaxies over the past fluid worldline (blue line in \figs{geometry_1D}{geometry_2D}).

\begin{figure}
\centering
\includegraphics[width=0.85
\columnwidth]{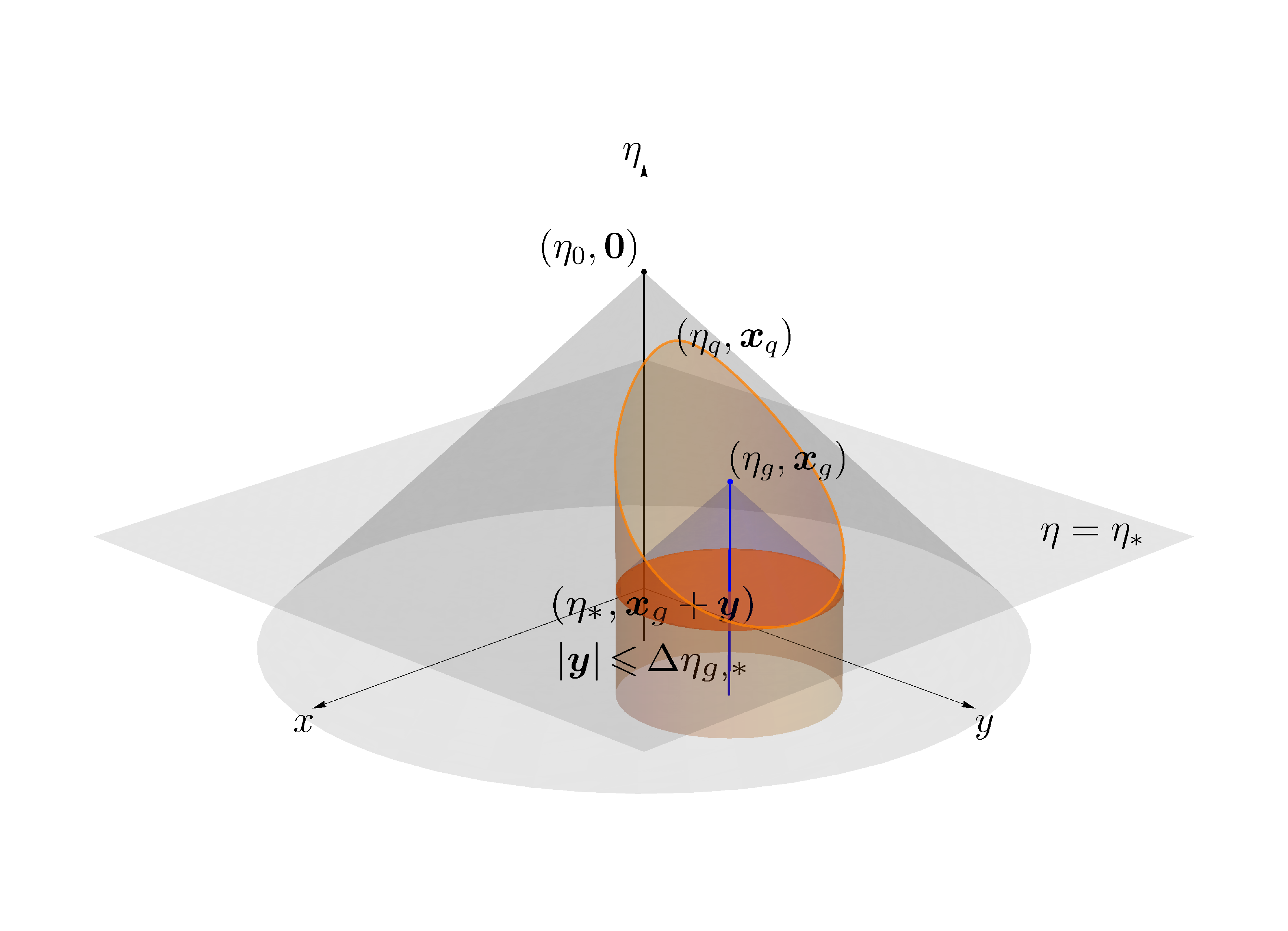} 
\caption{Same as \fig{geometry_1D}, but suppressing only one space dimension (with $\Delta \eta_{{g},\ast}$ defined as $\eta_{g}-\eta_\ast$). 
We see that there is only one position on the surface $\eta=\eta_\ast$ 
from which we could in principle receive the radiation directly (without it being scattered towards us at $(\eta_{g},\vec{x}_{g})$): 
however this point is hidden behind the observed galaxies.} 
\label{fig:geometry_2D}
\end{figure}

At leading order in $\Delta\eta\ll1/{\cal H}(\eta_\ast)$, and in absence of scattering and of additional sources after $\eta_\ast$, 
the radiation intensity received by the galaxies at some event $(\eta,\vec{x})$ on $x^\mu_{\rm fl}$ takes the form (see Appendix \ref{app:solution_single_flash} for a derivation) 
\begin{equation}
\begin{split}
\label{eq:observed_intensity}
{\cal I}(\eta,\vec{x},E,\vers{n}) &= \bigg(\frac{1+z(\eta)}{1+z_\ast}\bigg)^3e^{-\tau(\eta,\,\vec{x},\,E,\,\vers{n})}\, 
{\cal I}_\ast\big(\eta_\ast,\vec{x}+\vers{n}(\eta-\eta_\ast),E(\eta_\ast,\eta)\big) \\
&= \bigg(\frac{1+z(\eta)}{1+z_\ast}\bigg)^3e^{-\tau(\eta,\,\vec{x},\,E,\,\vers{n})}\,\frac{\Delta\eta\,a_\ast\,{\rho}_{\rm em}\big(\eta_\ast,\vec{x}+\vers{n}(\eta-\eta_\ast)\big)\,
\varepsilon_{\rm em}\big(\eta_\ast,E(\eta_\ast,\eta)\big)}{4\pi}\,\,,
\end{split}
\end{equation}
where the emitted intensity ${\cal I}_\ast$ is equal to $\Delta\eta(\rho_{\rm em}\varepsilon_{\rm em}a)|_{\eta=\eta_\ast}/(4\pi)$ 
(and is independent of $\vers{n}$ given that $\varepsilon_{\rm em}$ is isotropic), 
and the redshift dependence of the photon energy is, for general $\eta'\leq\eta$, given by 
\begin{equation}
\label{eq:energy_redshift_relation}
E(\eta',\eta)=E\frac{1+z(\eta')}{1+z(\eta)}\,\,,
\end{equation}
so that $E$ is the energy measured by an observer at $(\eta,\vec{x})$. Given 
the number density of absorbers and the absorption cross section, the optical depth $\tau$ in \eq{observed_intensity} is equal to 
\begin{equation}
\label{eq:optical_depth}
\tau(\eta,\vec{x},E,\vers{n}) = \int_{\eta_\ast}^{\eta}\dif\eta'\,(\sigma_{\rm ab}n_{\rm ab} a)\big(\eta',\vec{x}+\vers{n}(\eta-\eta'),E(\eta',\eta)\big)\,\,.
\end{equation}
From now on we take $\sigma_{\rm ab}$ to be the bound-free cross section $\sigma_{\rm bf}$ for photoionization of the $1s$ state, 
and $n_{\rm ab}=n_\HI$ to be the corresponding number density of neutral hydrogen in the $1s$ state. 
We will discuss what happens if we consider the full cross section and the more general scenario of multiple absorbers 
(like higher levels of hydrogen or helium, for example) in Section \ref{ssec:caveats}. 
Moreover, since in this section we assume the absorbers to be homogeneous, $n_\HI = \widebar{n}_\HI$, $\tau$ does not depend on $\vers{n}$. 
Also, for simplicity of notation we will drop the bar {on $\widebar{n}_\HI$.} 

Let us then assume that there are some local properties of galaxies that depend on the received flux along their whole past history: 
for example, these could be the heating and cooling rates of the gas accreting onto the parent halo, 
which are in turn related to the star-formation rate. 
Therefore, we expect this dependence to be inherited by the galaxy number density $n_{g}$ \cite{Schmidt:2017lqe}. 
More precisely, we can parameterize its response to the intensity of ionizing radiation by means of a Green's function $G_{g}$ along the fluid worldline 
(the subscript ``$g$'' is to differentiate it from the Green's function for the $q$ tracers, that is discussed in Section \ref{ssec:q_tracers}). 
Such function encodes the ``UV physics'' of galaxy formation that the bias expansion is oblivious about. 
At zeroth order, the RT effects on the mean galaxy density can then be written as 
\begin{equation}
\label{eq:definition_of_G_g-A-1}
\begin{split}
{\widebar{n}_{g}|_{\rm ion}(\eta) } = 4\pi
\int_{\eta_\ast}^\eta\dif\eta'\int_0^{+\infty}\dif E\,{G}^{(0)}_{g}(\eta,\eta',E)\,\widebar{{\cal I}}(\eta',E)\,\,, 
\end{split}
\end{equation}
where $G_g^{(0)}$ is the zeroth-order Green's function and we have used the fact that $\widebar{{\cal I}}$ does not depend on $\vers{n}$. 
Similarly, at first order in perturbations, we have 
\begin{equation}
\label{eq:definition_of_G_g-A-2}
\begin{split}
&\delta{n}_{g}|_{\rm ion}(\eta,\vec{x}) = \int_{\eta_\ast}^\eta\dif\eta'\int\dif\vers{n}\int_0^{+\infty}\dif E\,{G}^{(1)}_{g}(\eta,\eta',E,\vers{n})\,
\delta{{\cal I}}\big(\eta',\vec{x}_{\rm fl}(\eta'),E,\vers{n}\big) \\
&\hphantom{\delta{n}_{g}|_{\rm ion}(\eta,\vec{x}) } = 
\int_{\eta_\ast}^\eta\dif\eta'\int\dif\vers{n}\int_0^{+\infty}\dif E\,{G}^{(1)}_{g}(\eta,\eta',E,\vers{n})\,\delta{{\cal I}}(\eta',\vec{x},E,\vers{n})\,\,, 
\end{split}
\end{equation}
where again we have used the fact that $\vec{x}_{\rm fl}$ is equal to $\vec{x}$ at the order we are working at. 

Before proceeding, let us discuss \eqsII{definition_of_G_g-A-1}{definition_of_G_g-A-2} in more detail. 
What we have done is to isolate the contribution of ionizing radiation to the evolution of the galaxy number density 
(hence the subscript ``$\rm ion$'' on $\widebar{n}_g$ and $\delta{n}_{g}$). 
At the order in perturbations we are working at, there is no loss of generality if we just sum this contribution to that arising from purely gravitational evolution, 
whose nonlocality scale is the halo Lagrangian radius $R(M_{h})$. That is, we can write the total dimensionless perturbation $\delta_g$ to the 
galaxy number density as the sum $\delta_g|_{\rm grav}+\delta_g|_{\rm ion}$, 
where $\delta_g|_{\rm ion}$ is obtained from \eqsII{definition_of_G_g-A-1}{definition_of_G_g-A-2} by simply taking the ratio $\delta n_g|_{\rm ion}/\widebar{n}_g|_{\rm ion}$. 
We also emphasize that we allowed for two different Green's functions, one for the response of the average galaxy number density and one for its perturbation. 
Indeed, there is no physical reason why the response should be the same at all orders in perturbations \cite{Desjacques:2016bnm}. 
Besides, as we discussed at the beginning of this section, the sources of ionizing radiation are also 
expected to be a biased tracer of the underlying matter distribution. Therefore, 
$\delta{\cal I}$ will have both a deterministic and a stochastic contribution. 
If these two components have different emission spectra (which also effectively leads to them having different mean free paths), 
it is possible for galaxies to have a different response to each of them: we will not consider this case in the following, 
and just comment briefly in Section \ref{ssec:galaxy_statistics} on how this would modify the results we obtain here. 

What are then the properties of these Green's functions, ${G_g^{(0)}}$ and ${G_g^{(1)}}$? 
Let us first focus on their dependence on the photon four-momentum. 
At this order in perturbations we do not expect the galaxy response to depend on the photon arrival direction $\vers{n}$: 
from the point of view of the large-scale bias expansion, this would amount to the presence of a preferred direction in the rest frame of the observer comoving with the fluid, 
which is forbidden by rotational invariance.\footnote{At higher orders in perturbations we can, for example, use vectors constructed from gradients 
of the matter overdensity.} Then, in order to make progress, we assume that the time and energy dependencies can be factorized. 
Moreover, we assume that the energy dependence of both Green's functions is proportional to the absorption cross section itself, 
\ie~we write 
\begin{equation}
\label{eq:definition_of_G_g-B}
G^{(i)}_g(\eta,\eta',E)={\cal G}\sigma_{\rm bf}(E)G^{(i)}_g(\eta,\eta')\quad\text{for $i=0,1$}\,\,,
\end{equation}
where ${\cal G}$ is a constant with dimensions of an inverse length squared. This assumption is justified 
if we are thinking of the effect of photoevaporation of the gas accreting onto the halo 
(in any case, our conclusions will not change even if we consider a more general dependence on $E$, as we will see in Section \ref{ssec:caveats}). 

We now turn to the time dependence of the Green's functions, to have an idea of what are the different scales involved in the problem. 
Clearly, in the case of an instantaneous response of galaxy formation to the incoming flux, 
they would both be a function of $\eta$ times $\delta(\eta-\eta')$. 
In general, there is a finite response time, \ie~we can imagine that 
the Green's functions are peaked around $\eta'=\eta$, with some finite width for the peak. 
The incoming ionizing radiation is (obviously) not interacting with the dark matter in the host halo, 
but with the gas accreting onto it. The time scale of this interaction is set by atomic physics, and then we expect it to be much faster 
than a Hubble time. Nevertheless, we also expect the star-formation rate to be tied to the accretion rate of baryonic matter on the host halo. 
This, in turn, is related to the total mass flow, which we know from gravity-only $N$-body simulations to be controlled by Hubble. 
For this reason, in this section we are going to consider the case in which the galaxy response is varying only on cosmological time scales. 
A more detailed discussion is presented in Section \ref{sec:conclusions}. 
With this assumption, there is no loss of generality in redefining the two Green's functions 
in such a way as to reabsorb the factor $(1+z(\eta'))^3/(1+z_\ast)^3$ in \eq{observed_intensity}. 
More precisely, we write
\begin{equation}
\label{eq:reabsorb_redshift}
G^{(i)}_{g}(\eta,\eta')\to\bigg(\frac{1+z_\ast}{1+z(\eta')}\bigg)^3\,{G}^{(i)}_{g}(\eta,\eta')\quad\text{for $i=0,1$}\,\,.
\end{equation}
We will later use the fact that the responses vary on time scales of order Hubble. 
We can now carry out the integral over $\dif\eta'$ and $\dif\vers{n}$ in \eq{definition_of_G_g-A-2}. 
The full calculation is carried out in Appendix \ref{app:solution_single_flash}: 
eventually, we see that $\delta n_g|_{\rm ion}$ is equal to 
\begin{equation}
\label{eq:J_ion}
\begin{split}
&\delta n_g|_{\rm ion}(\eta,\vec{x})=\frac{\Delta\eta\,a_\ast\,{\cal G}}{4\pi}\int_{\abs{\vec{y}}\,\leq\,\eta\,-\,\eta_\ast}\dif^3y\, 
\frac{{G}^{(1)}_{g}(\eta,\eta_\ast+\abs{\vec{y}})}{\abs{\vec{y}}^2}\,\delta\rho_{\rm em}(\eta_\ast,\vec{x}+\vec{y})\,\times \\
&\hphantom{\delta n_g|_{\rm ion}(\eta,\vec{x})=\frac{\Delta\eta\,a_\ast\,{\cal G}}{4\pi} } 
\int_0^{+\infty}\dif E\,\sigma_{\rm bf}(E)\,\varepsilon_{\rm em}\big(\eta_\ast,E(\eta_\ast,\eta_\ast+\abs{\vec{y}})\big)\,{e^{-\hat{\tau}(\eta,\,\vec{y},\,E)}}\,\,,
\end{split}
\end{equation}
where $\hat{\tau}$ is given by 
\begin{equation}
\label{eq:hat_tau}
\hat{\tau}(\eta,\vec{y},E) = \hat{\tau}(\eta,\abs{\vec{y}},E) = \abs{\vec{y}}\int_{0}^{1}\dif u\,(n_\HI a)(\eta_\ast+u\abs{\vec{y}})\,
\sigma_{\rm bf}\big(E(\eta_\ast+u\abs{\vec{y}},\eta_\ast+\abs{\vec{y}})\big)\,\,.
\end{equation}

The photoionization of $1s$ hydrogen requires energies higher than the Lyman limit $E_{\infty}$: 
therefore the integral over $E$ in \eq{J_ion} starts from $E_{\infty}$.\footnote{Notice that 
$E(\eta',\eta)\geq E$ for $\eta'<\eta$ (photons lose energy as they travel towards the galaxy): therefore this constraint is automatically satisfied in \eq{hat_tau}.} 
At energies much higher than $E_\infty$, but small enough that the emitted electron is still nonrelativistic, 
$\sigma_{\rm bf}(E)\sim E^{-7/2}$ (see \eg~\cite{Sakurai:1167961,Rybicki:2004hfl}). More precisely, we have
\begin{equation}
\label{eq:photoionization_cross_section_leading_scaling}
\sigma_{\rm bf}(E) = \frac{64\pi\alpha^3}{3E_\infty^2}\bigg(\frac{E_\infty}{E}\bigg)^{\frac{7}{2}}\equiv\sigma_\infty\bigg(\frac{E_\infty}{E}\bigg)^{\frac{7}{2}}\,\,,
\end{equation}
where $\alpha=e^2/4\pi$ is the fine-structure constant. 
Corrections to this scaling can be expanded as a series in $E_\infty/E$, and are discussed in more detail in Section \ref{ssec:caveats}. 
Let us then make three further assumptions. First, we consider a power-law 
spectrum of emitted radiation, with a spectral index $s$ (which we take to be not too hard, \ie~$s < 1$): this is, for example, the parameterization 
discussed in \cite{Zhang:2006kr,DAloisio:2013mgn}, while \cite{Mao:2014rja} studies the cases $s=0$ and $s=-2$. 
Therefore, in \eq{J_ion} we can write $\varepsilon_{\rm em}(\eta_\ast,E)$ as $\varepsilon_{\infty}(\eta_\ast)(E/E_\infty)^{s}$. 
Finally, we further approximate $\hat{\tau}(\eta,\abs{\vec{y}},E)$ as 
\begin{equation}
\label{eq:hat_tau_approx}
\begin{split}
\hat{\tau}(\eta,\abs{\vec{y}},E) &= 
\abs{\vec{y}}\int_{0}^{1}\dif u\,(n_\HI a)(\eta_\ast+u\abs{\vec{y}})\,
\sigma_{\rm bf}\big(E(\eta_\ast+u\abs{\vec{y}},\eta_\ast+\abs{\vec{y}})\big) \\
&\approx\abs{\vec{y}}\sigma_{\rm bf}\big(E(\eta_\ast,\eta_\ast)\big)(n_\HI a)(\eta_\ast) 
=\abs{\vec{y}}\sigma_{\rm bf}(E)(n_\HI a)(\eta_\ast) \equiv \frac{\abs{\vec{y}}}{\lambda_{\rm ion}(\eta_\ast,E)}\,\,.
\end{split}
\end{equation}
Clearly, the time evolution of the mean number density of neutral hydrogen must be taken into account, 
since it enters in \eq{hat_tau_approx} and gives an additional dependence on $\abs{\vec{y}}$. 
We know that the ionization fraction basically changes from $0$ to $1$ during reionization, 
and it is possible that this change happened very quickly, leading to a strong dependence of $\hat{\tau}$ on $\abs{\vec{y}}$ 
(however, it is important to stress that we do not have yet a clear picture of the time evolution of the ionization fraction, 
since the main constraints come from the measurement of CMB anisotropies, that are only sensitive to the optical depth to redshift $z\approx 1100$). 
Therefore it is important to take this time dependence into account as well: 
we will come back to this point in Section \ref{ssec:caveats}, together with the full dependence of the cross section on redshift in \eq{hat_tau_approx}. 
For now, the approximation of \eq{hat_tau_approx} is sufficient. With these assumptions, then, \eq{J_ion} becomes
\begin{equation}
\label{eq:J_ion_approximated}
\begin{split}
&\delta n_g|_{\rm ion}(\eta,\vec{x})=\frac{\Delta\eta\,a_\ast
\,{\cal G}\,\sigma_\infty\,\varepsilon_\infty}{4\pi}\int_{\abs{\vec{y}}\,\leq\,\eta\,-\,\eta_\ast}\dif^3y\, 
\frac{{G}^{(1)}_{g}(\eta,\eta_\ast+\abs{\vec{y}})}{\abs{\vec{y}}^2}\,\bigg(\frac{1+z_\ast}{1+z(\eta_\ast+\abs{\vec{y}})}\bigg)^s\,\times \\ 
&\hphantom{\delta n_g|_{\rm ion}(\eta,\vec{x})=\frac{\Delta\eta\,a_\ast
\,{\cal G}\,\sigma_\infty\,\varepsilon_\infty}{4\pi} } 
\,\delta\rho_{\rm em}(\eta_\ast,\vec{x}+\vec{y})
\int_{E_\infty}^{+\infty}\dif E\,\bigg(\frac{E}{E_\infty}\bigg)^{s\,-\,\frac{7}{2}}\,{e^{-\frac{\abs{\vec{y}}}{\lambda_{\rm ion}(\eta_\ast,\,E)}}}\,\,. 
\end{split}
\end{equation}

In the following we are going to neglect the restriction of the integral to the interior of the past light cone of $(\eta,\vec{x})$, 
since the mean free path is much shorter than $\eta-\eta_\ast$. 
Indeed, if we imagine to observe galaxies at $z=\num{1.5}$, which is a typical redshift for the upcoming large-scale galaxy redshift surveys, 
we have that the ratio between $(\eta-\eta_\ast)$ and the m.f.p.~is of order $100$, even taking conservatively $\eta_\ast$ corresponding to the end of reionization ($z\approx 6$), 
and using that the m.f.p.~is of order $\lmpch{50}$ around these redshifts \cite{Worseck:2014fya,Becker:2014oga}. 
In any case, including the constraint $\abs{\vec{y}}\leq\eta-\eta_\ast$ is straightforward, but it would only complicate the calculations without adding relevant physics: 
we briefly discuss it in Section \ref{ssec:caveats}. Moreover, similarly to what we did in \eq{reabsorb_redshift} and without loss of generality, we reabsorb 
the factor $(1+z_\ast)^s/(1+z(\eta_\ast+\abs{\vec{y}}))^s$ coming from the redshift dependence of the emissivity into the Green's function. 
The integral over energy in \eq{J_ion_approximated} can then be carried out analytically. Let us however \mbox{approximate it by} 
\begin{equation}
\label{eq:integral_over_energy}
\int_{E_\infty}^{+\infty}\dif E\,\bigg(\frac{E}{E_\infty}\bigg)^{s\,-\,\frac{7}{2}}\,{e^{-\frac{\abs{\vec{y}}}{\lambda_{\rm ion}(\eta_\ast,\,E)}}} \approx 
\frac{2E_{\infty}}{5-2s}\,
e^{{-\frac{2s\,-\,5}{2(s\,-\,6)}}\abs{\vec{y}}\sigma_\infty(n_\HI a)(\eta_\ast)}\,\,,
\end{equation}
from which we can define an ``effective'' m.f.p.~${\lambda}_{\rm eff}(\eta_\ast)$ such that the integral is proportional to 
$\exp({-\abs{\vec{y}}}/{\lambda}_{\rm eff}(\eta_\ast))$. This approximation works well for $\abs{\vec{y}}\ll{\lambda}_{\rm eff}(\eta_\ast)$, 
with corrections starting at order $\abs{\vec{y}}^2/{\lambda}^2_{\rm eff}(\eta_\ast)$ (we will discuss in Section \ref{ssec:caveats} how to go beyond this approximation). 
To summarize, we have reduced \eq{J_ion_approximated} to the form 
\begin{equation}
\label{eq:J_ion_more_approximated}
\begin{split}
\delta n_g|_{\rm ion}(\eta,\vec{x})&=\underbrace{\frac{\Delta\eta\,a_\ast
\,{\cal G}\,\sigma_\infty\,\varepsilon_\infty\,E_\infty}{2\pi(5-2s)}}_{
\hphantom{{\cal C}_{\rm ion}\,}\equiv\,{\cal C}_{\rm ion}}\int\dif^3y\, 
\frac{{G}^{(1)}_{\rm g}(\eta,\eta_\ast+\abs{\vec{y}})}{\abs{\vec{y}}^2}\,\delta\rho_{\rm em}(\eta_\ast,\vec{x}+\vec{y})\,
{e^{-\frac{\abs{\vec{y}}}{{\lambda}_{\rm eff}(\eta_\ast)}}}\,\,.
\end{split}
\end{equation}
For simplicity of notation we will regroup the overall factors in a single dimensionless quantity ${\cal C}_{\rm ion}$ in the following. 
Moreover, for the rest of this section we will always intend $\lambda_{\rm eff}$ as evaluated at $\eta_\ast$ 
and $\cal H$ as evaluated at $\eta$, unless stated otherwise. 
We now use the fact that the responses vary on time scales of order Hubble by expanding $G^{(1)}_g$ in a Taylor series around $\eta_\ast$:\footnote{We could 
have equivalently expanded the Green's function around $\eta$. We have chosen $\eta_\ast$ as the expansion point since this makes calculations easier.} 
\begin{equation}
\label{eq:G_g_expansion}
G^{(1)}_{g}(\eta,\eta') = \sum_{n\,=\,0}^{+\infty}g_{n}(\eta){\cal H}^n(\eta)(\eta'-\eta_\ast)^n\,\,,
\end{equation}
where, without loss of generality, we have made the functions $g_{n}(\eta)$ dimensionless by factoring out ${\cal H}^n(\eta)$. 
The response for ${\widebar{n}_{g}|_{\rm ion}(\eta)}$ can be expanded in a similar way, but this is unnecessary (as we will see in a moment). 

Introducing the dimensionless perturbation $\delta_{\rm em}=\delta\rho_{\rm em}/\widebar{\rho}_{\rm em}$ of the number density of emitters, \eq{J_ion_more_approximated} becomes
\begin{equation}
\label{eq:J_ion_more_approximated-bis}
\begin{split}
\delta n_g|_{\rm ion}(\eta,\vec{x})
&={\cal C}_{\rm ion}\widebar{\rho}(\eta_\ast)\sum_{n\,=\,0}^{+\infty}g_{n}(\eta){\cal H}^{n}(\eta)\int\dif^3y\,
\abs{\vec{y}}^{n\,-\,2}\,\delta_{\rm em}(\eta_\ast,\vec{x}+\vec{y})\,{e^{-\frac{\abs{\vec{y}}}{{\lambda}_{\rm eff}(\eta_\ast)}}}\,\,, 
\end{split}
\end{equation}
Finally, after dividing \eq{J_ion_more_approximated-bis} 
by the average galaxy number density $\widebar{n}_g|_{\rm ion}$ we obtain the expression for the dimensionless fluctuation $\delta_g|_{\rm ion}$, \ie~ 
\begin{equation}
\label{eq:flux_perturbations}
\delta_g|_{\rm ion}(\eta,\vec{x})=\frac{{\cal C}_{\rm ion}\widebar{\rho}(\eta_\ast)}{\widebar{n}_g|_{\rm ion}(\eta)}\sum_{n\,=\,0}^{+\infty}g_{n}(\eta){\cal H}^{n}\int\dif^3y\,
\abs{\vec{y}}^{n\,-\,2}\,\delta_{\rm em}(\eta_\ast,\vec{x}+\vec{y})\,{e^{-\frac{\abs{\vec{y}}}{{\lambda}_{\rm eff}}}}\,\,.
\end{equation}

\begin{figure}
\centering
\includegraphics[width=0.85\columnwidth]{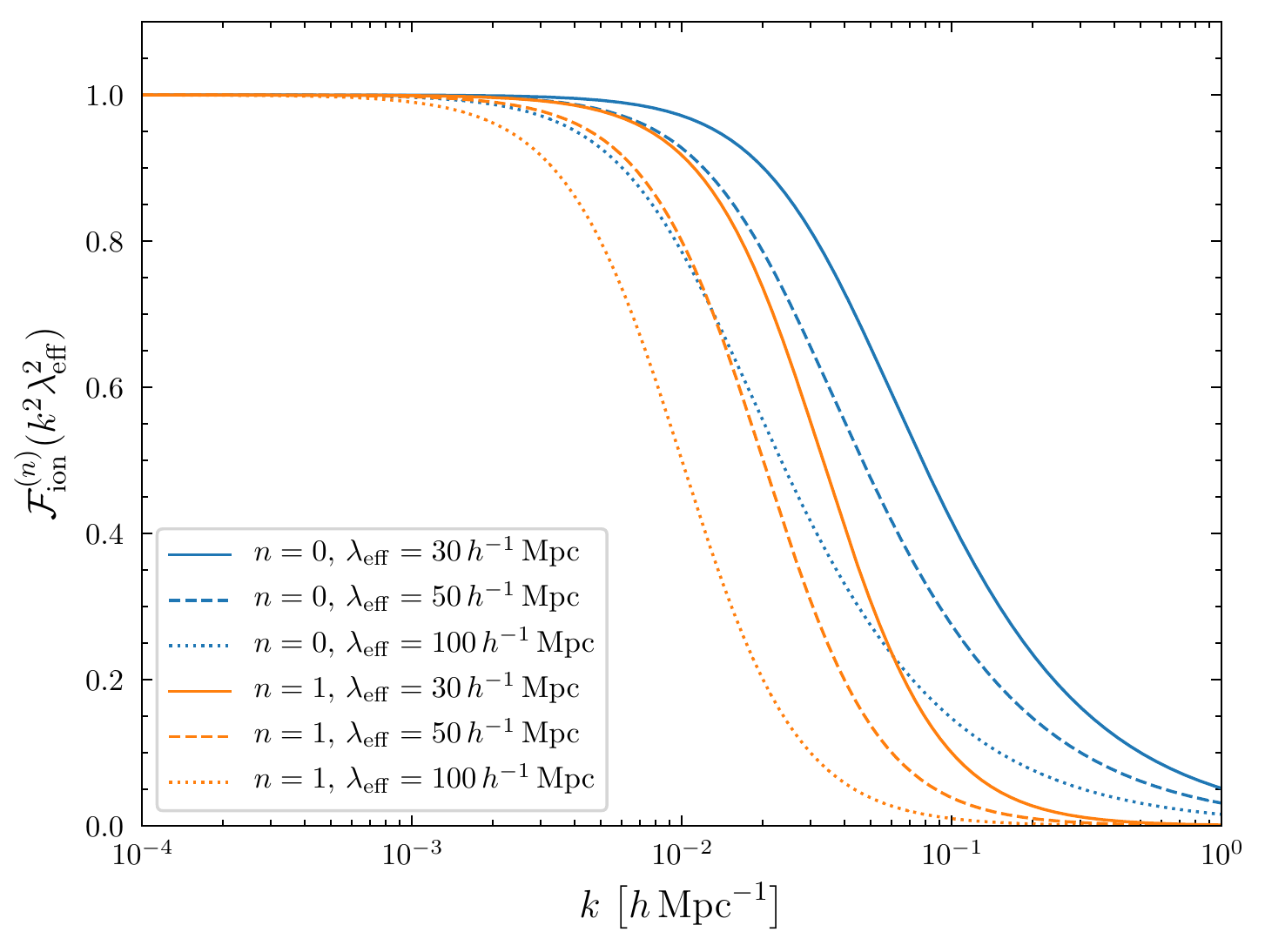} 
\captionsetup{singlelinecheck=off}
\caption[.]{Plot of the first two functions ${\cal F}^{(n)}_{\rm ion}$ for different values of $\lambda_{\rm eff}$. At large $k\gg 1/\lambda_{\rm eff}$ 
they decay as $1/k^{n\,+\,1}$, so that their different scale dependence can in principle \mbox{be distinguished.}} 
\label{fig:PS_corrections}
\end{figure}

What does \eq{flux_perturbations} mean for the bias expansion of $\delta_{g}|_{\rm ion}$? We 
see that \eq{flux_perturbations} is an example of a nonlocal contribution to the bias expansion: 
only if its kernel is sufficiently localized is it possible to do a (spatial) derivative expansion and end up with local operators. 
Here it is the m.f.p.~that controls the spatial extent of the kernel (galaxies are sensitive 
only to ionizing photons coming from a comoving volume of size $\sim\lambda_{\rm eff}^3$ along the past fluid trajectory), 
so the expansion will be in $\lambda_{\rm eff}\vec{\nabla}$. 
Consequently, all the higher-derivative operators become important at momenta $k\sim1/\lambda_{\rm eff}$. 
This is the conclusion obtained in \cite{Schmidt:2017lqe}. However, we also see that \eq{flux_perturbations} is actually a resummation of all these higher-derivative terms: 
if we can treat the sum over $n$ perturbatively, we are able to predict the scale dependence of $\delta_g|_{\rm ion}$ 
also for $k\gtrsim 1/\lambda_{\rm eff}$ in terms of a finite number of functions of time. 
This can be seen more easily if we work in Fourier space. \eq{flux_perturbations} is a convolution in real space, so $\delta_g|_{\rm ion}(\eta,\vec{k})$ is 
\begin{equation}
\label{eq:delta_perturbations-A}
\delta_g|_{\rm ion}(\eta,\vec{k}) = f_{\rm ion}(\eta,k^2)\delta_{\rm em}(\eta_\ast,\vec{k})\,\,,
\end{equation}
where $f_{\rm ion}(\eta,k^2)$ (which can depend only on $k$ because the kernel in \eq{flux_perturbations} depends only on $\abs{\vec{y}}$, 
and cannot contain odd terms in $k$ because of locality) is given by 
\begin{equation}
\label{eq:delta_perturbations-B}
f_{\rm ion}(\eta,k^2) = \frac{{\cal C}_{\rm ion}\widebar{\rho}(\eta_\ast)}{\widebar{n}_g|_{\rm ion}(\eta)}
\sum_{n\,=\,0}^{\infty}g_{n}(\eta){\cal H}^{n}\int\dif^3y\,e^{{-}i\vec{k}\cdot\vec{y}}\,\abs{\vec{y}}^{n\,-\,2}\,
{e^{-\frac{\abs{\vec{y}}}{{\lambda}_{\rm eff}}}}\,\,. 
\end{equation}
We can simplify this expression for $f_{\rm ion}$ if we define 
\begin{equation}
\label{eq:redefine_f_ion-A}
\int\dif^3y\,e^{{-}i\vec{k}\cdot\vec{y}}\,\abs{\vec{y}}^{n\,-\,2}\,
{e^{-\frac{\abs{\vec{y}}}{{\lambda}_{\rm eff}}}}\equiv\lambda_{\rm eff}^{n\,+\,1}\,(4\pi n!)\,{\cal F}^{(n)}_{\rm ion}(k^2\lambda^2_{\rm eff})\,\,, 
\end{equation}
where ${\cal F}^{(n)}_{\rm ion}$ are dimensionless functions of $k^2\lambda^2_{\rm eff}$, 
and the overall factor of $4\pi n!$ has been chosen so that they are all equal to $1$ for $k=0$. 
Further, since $\widebar{n}_g|_{\rm ion}(\eta)$ depends only on time, for each $n$ we can reabsorb the dimensionless factor 
$(4\pi n!)\times (\lambda_{\rm eff}{\cal C}_{\rm ion}\widebar{\rho}(\eta_\ast))/(\widebar{n}_g|_{\rm ion}(\eta))$ 
into the corresponding coefficient $g_n$. With these redefinitions, $f_{\rm ion}$ takes the form
\begin{equation}
\label{eq:redefine_f_ion-B}
f_{\rm ion}(\eta,k^2)={\sum_{n\,=\,0}^{+\infty}g_{n}(\eta)({\cal H}\lambda_{\rm eff})^n {\cal F}^{(n)}_{\rm ion}(k^2\lambda^2_{\rm eff})}\,\,, 
\end{equation}
and in the following we will often suppress the time dependencies of the coefficients $g_{n}$ for simplicity of notation. 
For future reference, we also write down the explicit expression of the functions ${\cal F}^{(n)}_{\rm ion}$, \ie~ 
\begin{equation}
\label{eq:functions_f_n}
{\cal F}^{(n)}_{\rm ion}(x^2)=\frac{\sin\!\big(n\arctan(x)\big)}{nx(1+x^2)^{n/2}}\,\,, 
\end{equation}
with $f^{(0)}_{\rm ion}(x^2)=4\pi\arctan(x)/x$. A plot of the first two functions $\smash{{\cal F}^{(n)}_{\rm ion}}$ is shown in \fig{PS_corrections}.

As we discussed at the end of Section \ref{sec:questions}, we 
assume that $\delta_{\rm em}$ can be expressed in terms of the matter overdensity via a bias expansion 
with a nonlocality scale much shorter than the m.f.p.~of ionizing radiation. Therefore everything depends on whether or not we 
are able to truncate the sums in \eq{redefine_f_ion-B} at a finite order: otherwise we would need to know an infinite number of functions of $\eta$ 
(\ie~the $g_{n}$) to compute $f_{\rm ion}$. Fortunately, this is possible, since the m.f.p.~of ionizing radiation is much shorter than ${\cal H}^{-1}$. Therefore 
we can safely truncate the sums at some finite order in ${\cal H}{\lambda}_{\rm eff}$, 
and correspondingly introduce only a finite number of new bias coefficients in the bias expansion for galaxies.\footnote{We 
notice that in the limit of the effective mean free path going to zero there is no response of the galaxy number density to ionizing radiation if the 
Green's function for $\delta n_g|_{\rm ion}$ is zero at $\eta=\eta_\ast$. This makes sense since in this limit 
only radiation emitted from events along the fluid worldline could have reached the galaxies.} 

We emphasize again how \eq{redefine_f_ion-B} has nothing to do with an expansion of \eq{delta_perturbations-B} in powers of $k^2{\lambda}^2_{\rm eff}$. 
The point here is that the scale dependence at each order in ${\cal H}{\lambda}_{\rm eff}$ can be computed non-perturbatively 
in terms of the single parameter ${\lambda}_{\rm eff}$. By solving the full 
RT equation, and by simple locality arguments on the response of galaxies to the ionizing radiation, 
we have obtained a resummation of the expansion in spatial derivatives. This gives a precise correction to the shape of the power spectrum of galaxies 
with an amplitude set by a finite number of new bias coefficients, \ie~the $g_{n}$ appearing in \eq{redefine_f_ion-B}. 

Before proceeding to the next section, where we are going to compute the power spectrum of galaxies, 
we also recall that, given that we see a suppression of radiative-transfer effects in the clustering of high-mass galaxies \cite{Schmidt:2017lqe}, 
the Green's function for $\delta n_{\rm g}|_{\rm ion}$ must be suppressed by some factor $p(M_h,M_{\rm J})$ 
with respect to the one for the average galaxy density, with $p(M_h,M_{\rm J})\ll 1$ for $M_h\gg M_{\rm J}$. 
This is clear since, as we can see from \eqsII{definition_of_G_g-A-1}{delta_perturbations-B}, the effect of radiative transfer on the fractional galaxy density perturbation 
is controlled by the logarithmic response of the galaxy density to the radiation, that is the ratio of the two Green's functions in 
\eqsII{definition_of_G_g-A-1}{definition_of_G_g-A-2}.

\subsection{Galaxy statistics}
\label{ssec:galaxy_statistics}

\noindent We can now compute the contribution to the galaxy power spectrum from \eq{delta_perturbations-A}. 
This equation relates $\delta_g|_{\rm ion}$ at $\eta$ with $\delta_{\rm em}$ at $\eta_\ast$: that is, we only need to know the 
statistics of the sources at $\eta_\ast$ to compute the contribution of RT effects to $P_{gg}$ at $\eta$. 
At leading order in perturbations and derivatives we can write 
\begin{equation}
\label{eq:delta_em_bias_expansion-A}
\delta_{\rm em}(\eta,\vec{k}) = b_{\rm em}(\eta)\delta(\eta,\vec{k}) + \epsilon_{\rm em}(\eta,\vec{k})\,\,,
\end{equation}
where $b_{\rm em}$ is the linear LIMD bias, and the operator $\epsilon_{\rm em}$ (not to be confused with the emissivity) 
captures the effect of short-scale physics on the evolution of the sources at this order in perturbations (see \eg~Section 2 of \cite{Desjacques:2016bnm} for details). 
The stochastic term $\epsilon_{\rm em}$ is uncorrelated with $\delta$ and has a $k$-independent correlation function (up to terms of order $k^2R^2(M_h)$), \ie\footnote{Notice 
that, similarly to what happens to dark matter halos (or any other tracer), 
stochasticity is never completely uncorrelated with large-scale density fluctuations: 
gravitational evolution couples long- and short-wavelength modes, so that at higher orders in perturbations terms like $\epsilon_{{\rm em},\delta}\delta$ 
(where $\epsilon_{{\rm em},\delta}$ is a different operator than $\epsilon_{\rm em}$) must be included. See, \eg, \cite{Desjacques:2016bnm} for a detailed discussion.} 
\begin{equation}
\label{eq:delta_em_bias_expansion-B}
\braket{\epsilon_{\rm em}(\eta,\vec{k})\epsilon_{\rm em}(\eta',\vec{k}')}'=P_{\epsilon_{\rm em}}^{\{0\}}(\eta,\eta')\,\,.
\end{equation}

Without loss of generality, we can write $\delta_{\rm em}(\eta_\ast,\vec{k})$ in terms of $\delta(\eta,\vec{k})$ by 
reabsorbing the linear growth factor ${D_1(\eta_\ast)}/{D_1(\eta)}$ in the bias parameter $b_{\rm em}(\eta_\ast)$. 
Consequently, through $\delta_{g}|_{\rm ion}(\eta,\vec{k})$ the bias expansion for galaxies will gain the two new terms 
\begin{equation}
\label{eq:terms_in_bias_expansion}
\delta_{g}|_{\rm ion}(\eta,\vec{k}) = {f_{\rm ion}(\eta,k^2)}\big[b_{\rm em}(\eta_\ast)\delta(\eta,\vec{k}) + \epsilon_{\rm em}(\eta_\ast,\vec{k})\big]\,\,.
\end{equation}
We emphasize a minor point about this expression for $\delta_{g}|_{\rm ion}$, 
that will nevertheless become relevant when we consider multiple emission times in Section \ref{ssec:generalization}. 
Here $\delta_{g}|_{\rm ion}$ is a sum of a deterministic and a stochastic term, each multiplied by a function of $k^2$. 
Even if the overall amplitude of these two functions is different, their scale dependence is the same (\ie~it is given by the single function $f_{\rm ion}$). 
Moreover, we can also see what would happen if we considered two different Green's function for the stochastic and deterministic contributions 
to $\delta{\cal I}$ in \eq{definition_of_G_g-A-2}. The calculations leading to \eq{terms_in_bias_expansion} go through in 
the same way: the only difference is that instead of a single function $f_{\rm ion}$ we now have two different functions 
multiplying $\delta$ and $\epsilon_{\rm em}$. 
These two functions both have an expansion like that of \eq{redefine_f_ion-B}, but each of them has its own set of coefficients $g_{n}$: 
this captures the difference between the Green's functions for the stochastic and deterministic contributions. 

The full galaxy fractional overdensity is simply given by the sum of $\delta_{g}|_{\rm grav}$ and $\delta_{g}|_{\rm ion}$ at this order in perturbations, 
as we have seen at the beginning of Section \ref{ssec:single_flash}. That is, it takes the form 
\begin{equation}
\label{eq:galaxy_overdensity}
\begin{split}
\delta_g(\eta,\vec{k}) &=\delta_{g}|_{\rm grav}(\eta,\vec{k})+\delta_{g}|_{\rm ion}(\eta,\vec{k}) \\
&= b_1(\eta)\delta(\eta,\vec{k})+\epsilon_g(\eta,\vec{k}) 
+ {f_{\rm ion}(\eta,k^2)}\big[b_{\rm em}(\eta_\ast)\delta(\eta,\vec{k}) + \epsilon_{\rm em}(\eta_\ast,\vec{k})\big]\,\,,
\end{split}
\end{equation}
where $\delta_g|_{\rm grav}$ contains the linear LIMD and stochastic terms ($b_1\delta$ and $\epsilon_g$, respectively).

Let us then compute the equal-time galaxy power spectrum. For this purpose, it is useful to factor out 
$g_{0}$ from our expression for $f_{\rm ion}$, since in any case it is degenerate with the bias coefficient $b_{\rm em}$ 
and the amplitude of the stochastic term $\epsilon_{\rm em}$: that is, in \eq{galaxy_overdensity} we redefine
\begin{equation}
\label{eq:redefining_f_ion_to_simplify_P_gg-A}
f_{\rm ion}(\eta,k^2)\to{g_{0}(\eta)}\,f_{\rm ion}(\eta,k^2)\,\,,
\end{equation}
so that now we have
\begin{equation}
\label{eq:redefining_f_ion_to_simplify_P_gg-B}
f_{\rm ion}(\eta,k^2) = \mathcal{F}^{(0)}_{\rm ion}(k^2{\lambda}^2_{\rm eff}) + {\cal O}({\cal H}{\lambda}_{\rm eff})\,\,, 
\end{equation}
\ie~$f_{\rm ion}(\eta,k^2)$ is independent of the galaxy response to ionizing radiation at leading order in ${\cal H}\lambda_{\rm eff}$. 
If we further reabsorb $g_{0}$ into $b_{\rm em}$ and $\epsilon_{\rm em}$, the equal-time galaxy power spectrum $P_{gg}(\eta,k)$ takes the form 
\begin{equation}
\label{eq:galaxy_PS}
\begin{split}
P_{gg}(\eta,k) &= \big[b_1(\eta)+{b}_{\rm em}(\eta_\ast)f_{\rm ion}(\eta,k^2)\big]^2P_{\rm L}(\eta,k) + P^{\{0\}}_{\epsilon_{g}}(\eta) \\
&\;\;\;\; + 2f_{\rm ion}(\eta,k^2)P^{\{0\}}_{\epsilon_{g}\epsilon_{\rm em}}(\eta,\eta_\ast) + 
f^2_{\rm ion}(\eta,k)P^{\{0\}}_{\epsilon_{\rm em}}(\eta_\ast)\,\,,
\end{split}
\end{equation}
where $P_{\rm L}$ is the linear matter power spectrum and, following \cite{Desjacques:2016bnm}, 
we have defined the cross power spectrum between the two different stochastic terms $\epsilon_g$ and $\epsilon_{\rm em}$ as
\begin{equation}
\label{eq:stochastic_cross_PS}
\braket{\epsilon_{g}(\eta,\vec{k})\epsilon_{\rm em}(\eta',\vec{k}')}'\equiv P^{\{0\}}_{\epsilon_{g}\epsilon_{\rm em}}(\eta,\eta')\,\,.
\end{equation}
We notice that the contributions involving $\epsilon_{\rm em}$ are basically modifying the form of the higher-derivative corrections to the stochastic term $\epsilon_g$. 
These are present also if we consider gravitational interactions only: while in that case they become relevant at scales of order of the halo Lagrangian radius, 
here they are controlled by the m.f.p.~of ionizing radiation. 

From \eq{stochastic_cross_PS} we see that, if we stop at zeroth order in the expansion of \eq{redefining_f_ion_to_simplify_P_gg-B} 
(so that the function $f_{\rm ion}$ is univocally determined in terms of the mean free path), 
at fixed $\eta$ we can predict the galaxy power spectrum up to $4$ constants, 
\ie~$b_1$, ${b}_{\rm em}$, and the amplitudes of the two stochastic terms $\epsilon_g$, $\epsilon_{\rm em}$.\footnote{Of course we can consider also the 
effective mean free path as an additional free parameter, with a prior between $\sim\kmpch{30}$ and $\sim\kmpch{100}$, 
to account for uncertainties in the evolution of the mean density of neutral hydrogen.} 
We also see that there are no degeneracies between these four parameters. 
Clearly we can discriminate between $b_1,{b}_{\rm em}$ and $\smash{P^{\{0\}}_{\epsilon_g}},\smash{P^{\{0\}}_{\epsilon_{\rm em}}}$ thanks to the scale dependence of 
the matter power spectrum. Then, the scale dependence of $f_{\rm ion}$ allows to respectively distinguish $b_1$ from ${b}_{\rm em}$, 
and $\smash{P^{\{0\}}_{\epsilon_g}}$ from $\smash{P^{\{0\}}_{\epsilon_{\rm em}}}$. 

If we go beyond the zeroth order in ${\cal H}\lambda_{\rm eff}$, we need more free parameters to compute $f_{\rm ion}$. 
Thanks to the different scale dependence of the functions $\smash{{\cal F}^{(n)}_{\rm ion}}$, however, it is in principle possible to discriminate between them.

\begin{figure}
\centering
\includegraphics[width=0.85\columnwidth]{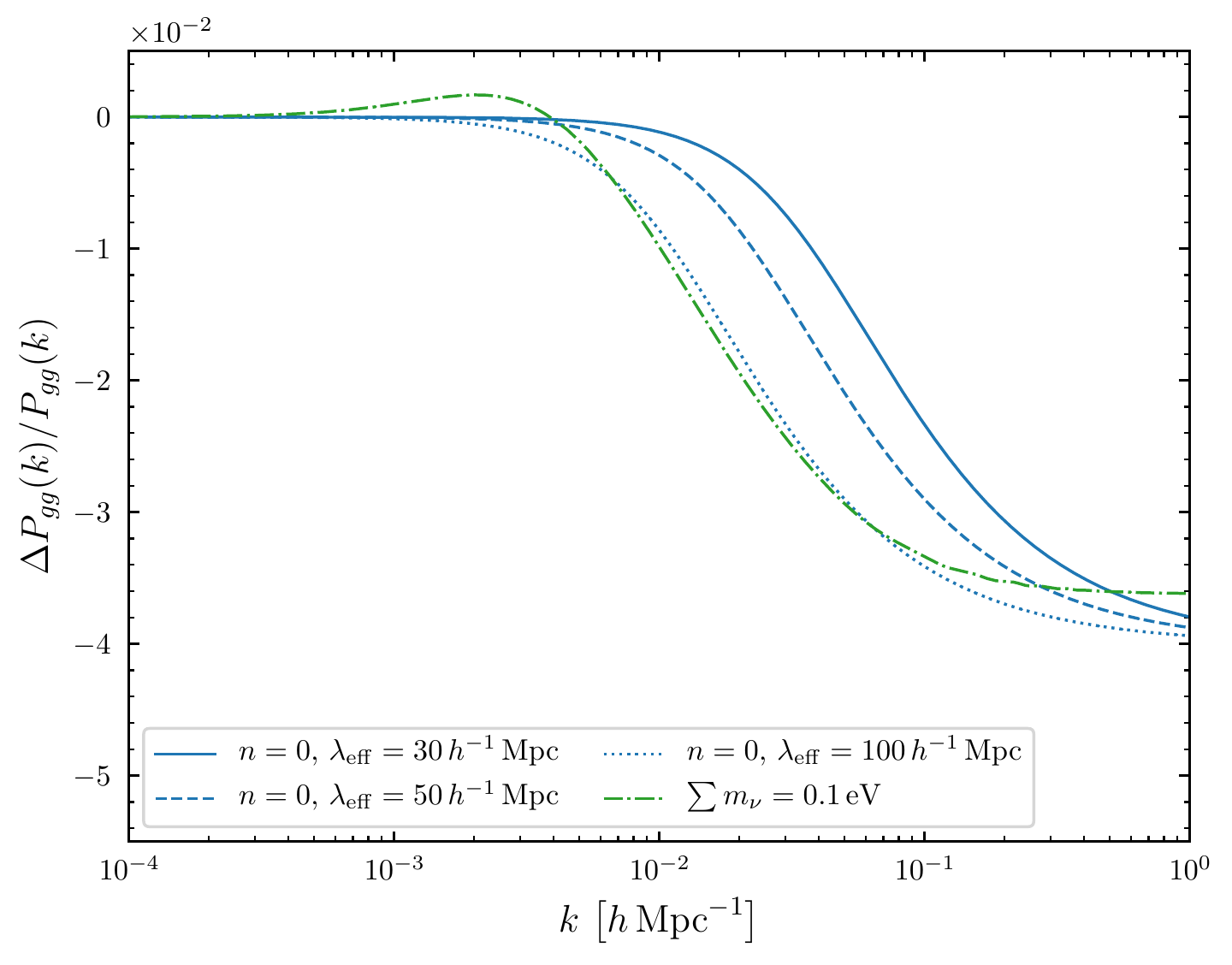} 
\captionsetup{singlelinecheck=off}
\caption[.]{Plot of $\smash{{\cal F}^{(0)}_{\rm ion}}$ for different values of $\lambda_{\rm eff}$ (blue curves). 
It has been multiplied by an RT bias coefficient (more precisely, by the combination $2b_{\rm em}(\eta_\ast)/b_1(\eta)$ of \eq{relative_difference}) 
with arbitrary value of $\num{0.04}$ (note that, depending on the parent halo mass and redshift, the RT bias coefficients could be small) 
and its limit for $k=0$ has been subtracted (since it can always be reabsorbed in the linear LIMD bias). 
The green dot-dashed curve shows the relative difference between the square of the transfer functions for the total matter overdensity (at $z=1$) 
with $\sum m_\nu = \SI{0.1}{\rm eV}$ and $m_\nu=0$.} 
\label{fig:neutrino_scale_dependence}
\end{figure}

We conclude this section with a brief discussion about the possible degeneracy between these RT effects and massive neutrinos. 
If we neglect the fact that massive neutrinos in general lead to a scale-dependent bias 
(see \cite{Ichiki:2011ue,Castorina:2013wga,LoVerde:2014pxa,Villaescusa-Navarro:2017mfx,Chiang:2017vuk,Chiang:2018laa}), the effect 
on the galaxy power spectrum is captured by the modification to the transfer function $T(k)$ for the total matter overdensity. More precisely, the relative correction to the 
deterministic part of the galaxy power spectrum is given by
\be
\frac{\Delta P_{gg}(\eta,k)}{P_{gg}(\eta,k)} = \frac{T^2(\eta,k, m_\nu\neq 0)}{T^2(\eta,k,m_\nu=0)}-1\,\,.
\ee
This is the green dot-dashed curve in \fig{neutrino_scale_dependence}. 
The corresponding correction due to RT effects, at leading order in $b_{\rm em}$, is given by (see \eq{galaxy_PS}) 
\begin{equation}
\label{eq:relative_difference}
\frac{\Delta P_{gg}(\eta,k)}{P_{gg}(\eta,k)} = \frac{2b_{\rm em}(\eta_\ast)}{b_1(\eta)}f_{\rm ion}(\eta,k^2)\,\,,
\end{equation}
where $f_{\rm ion}$ is given by \eq{redefining_f_ion_to_simplify_P_gg-B} and we recall that $b_{\rm em}$ must indeed be regarded as a small number 
since we reabsorbed the leading RT bias coefficient $g_{0}$ into it. Can this mimic the effect due to massive neutrinos? 
First, we see that the amplitude of this scale-dependent correction at $k=0$ is always degenerate with the linear LIMD bias. 
In other words, $b_1$ is fixed by the amplitude of $P_{gg}$ on large scales (obviously neglecting degeneracies with, \eg, $\sigma_8$, since 
they are not relevant for the sake of this discussion). This corresponds to redefining $b_1$ in \eq{galaxy_PS} as 
$b_1(\eta)\to b_1(\eta)-b_{\rm em}(\eta_\ast)f_{\rm ion}(\eta,0)$. 
Correspondingly \eq{relative_difference} sees simply $f_{\rm ion}(\eta,k^2)$ replaced by $f_{\rm ion}(\eta,k^2)-f_{\rm ion}(\eta,0)$. 
This is plotted in \fig{neutrino_scale_dependence} at leading order in ${\cal H}\lambda_{\rm eff}$ (blue curves). 
From the plot it is clear that these scale-dependent corrections are very similar in shape to those 
from massive neutrinos, \ie~RT effects could give rise to a bias in the constraints on $\sum m_\nu$ if not accounted for.\footnote{In 
principle we could think of getting around this degeneracy by isolating $f_{\rm ion}$ from the corrections to the stochasticity (see \eq{galaxy_PS}). 
However, we expect that neutrinos affect also the scale dependence of the stochastic term on scales $k\sim k_{\rm fs}$. 
Besides, as we are going to see in Section \ref{sec:full_RT}, when we consider the generic case of emission of radiation over Hubble time scales 
and add the inhomogeneities in the optical depth, the function of $k^2$ multiplying the stochastic term will not be the same as the one \mbox{multiplying $\delta$.}}

\subsection{Locally-biased tracers}
\label{ssec:q_tracers}

\noindent We now move to the $q$ tracers. These tracers are assumed to be locally sensitive to the ionizing radiation emitted by the sources. 
We capture their memory of the emitted radiation field by allowing their number density to depend on the integral over $\log E$ and $\vers{n}$ 
of the emission coefficient $(\rho_{\rm em}\varepsilon_{\rm em})/(4\pi)$, which is the total amount of ionizing photons emitted per unit time and unit volume. However, 
this is not enough. While we expect this dependence to be local in space, we must in principle integrate the emission coefficient also along the past fluid worldline, 
with a Green's function $G_{q}$ that describes how $n_{q}$ responds to the emissivity of the sources. As in the case of galaxies, we allow for two different Green's 
functions at zeroth and first order in perturbations. In principle these Green's functions can also depend on the photon energy 
(while they cannot depend on the photon direction $\vers{n}$ because of rotational invariance, 
at the order in perturbations we are working at). Therefore, we write $\delta {n}_q|_{\rm ion}$ as
\begin{equation}
\label{eq:q_tracers-A-2}
\begin{split}
\delta n_{q}|_{\rm ion}(\eta,\vec{x}) &= \int_0^\eta\dif\eta'\int_0^{+\infty}\frac{\dif E}{E}\int\dif\vers{n}\,G^{(1)}_{q}(\eta,\eta',E)\,
\frac{(\delta\rho_{\rm em}\varepsilon_{\rm em})(\eta',\vec{x},E)}{4\pi} \\ 
&= \int_0^\eta\dif\eta'\,\delta{\rho}_{\rm em}(\eta',\vec{x})\int_0^{+\infty}\frac{\dif E}{E}\,G^{(1)}_{q}(\eta,\eta',E)\,
\varepsilon_{\rm em}(\eta',E)\,\,, 
\end{split}
\end{equation}
where we have used our assumptions of an isotropic emissivity. A similar relation holds for $\widebar{n}_q|_{\rm ion}$. 
In \eq{q_tracers-A-2} we have also used the fact that, at the order we are working at, everything is comoving. In principle 
we should have written our integral over the emission coefficient as, schematically, 
$n_{q}\supset{-\int\rho_{\rm em}\varepsilon_{\rm em}U^\mu_{\rm em} U_\mu}$, where ${-\rho_{\rm em}}U^\mu_{\rm em}U_\mu$ is the energy 
density of the emitters as seen from the observer comoving with the fluid. 
Calling $\vec{v}_{\rm rel}$ the relative velocity between the $q$ tracers and the sources 
we have ${-U^\mu_{\rm em} U_\mu}\sim 1+\abs{\vec{v}_{\rm rel}}^2/{2}$, so that these effects are very suppressed in the nonrelativistic limit. 
Moreover, in presence of gravitational interactions only, $\vec{v}_{\rm rel}$ is only sourced starting from first order in derivatives, due to the equivalence principle. 

From \eq{q_tracers-A-2} we see that the integral in $\dif E$ can be redefined as a new Green's function that depends on $(\eta,\eta')$ only. 
In the case of instantaneous emission, then, the fractional overdensity $\delta_{q}|_{\rm ion}(\eta,\vec{x})$ is simply proportional 
to $\delta_{\rm em}(\eta_\ast,\vec{x})$ through a $\eta$-dependent function 
(even if this assumption is relaxed, using the bias expansion for $\delta_{\rm em}$ of \eq{delta_em_bias_expansion-A} 
we can still simply absorb the integral over $\eta'$ in the time dependence of the bias coefficients and stochastic terms, similarly 
to what happened when we moved from \eq{locality-B} to \eq{locality-C}). Therefore, the final expression for $\delta_{q}$ in Fourier space is simply 
\begin{equation}
\label{eq:q_tracers-B}
\delta_{q}(\eta,\vec{k}) = b_{q}(\eta)\delta(\eta,\vec{k}) + \epsilon_{q}(\eta,\vec{k}) + b_{\epsilon_{\rm em}}(\eta_\ast)\epsilon_{\rm em}(\eta_\ast,\vec{k})\,\,,
\end{equation}
where we kept the dependence on both stochastic terms $\epsilon_{q}$ and $\epsilon_{\rm em}$ (since they are different fields), 
while $b_{\rm em}\delta\subset\delta_{\rm em}$ was absorbed in the LIMD bias. 
We see that we cannot reabsorb the factor $b_{\epsilon_{\rm em}}$ in the amplitude of the stochastic term $\epsilon_{\rm em}$ 
without altering our expression for the galaxy power spectrum of \eq{galaxy_PS}. 

Before proceeding we notice that, as we discussed below \eq{definition_of_G_g-A-2}, 
also in this case we could allow for two different Green's function in \eq{q_tracers-A-2}, one for 
the response to the deterministic part of the emission coefficient and one for its stochastic part. 
It is straightforward to see, however, that \eq{q_tracers-B} would not change.

\subsection{\texorpdfstring{$gq$}{gq} cross-correlation}
\label{ssec:power_spectra}

\noindent It is straightforward to compute $P_{gq}$ and $P_{qq}$ using \eqsIII{galaxy_overdensity}{redefining_f_ion_to_simplify_P_gg-A}{q_tracers-B}: 
we find that $P_{gq}$ is equal to 
\begin{equation}
\label{eq:P_gq}
\begin{split}
P_{gq}(\eta,k) &= b_{q}(\eta)\big[b_1(\eta)+{b}_{\rm em}(\eta_\ast)f_{\rm ion}(\eta,k^2)\big]P_{\rm L}(\eta,k) + P^{\{0\}}_{\epsilon_{g}\epsilon_{q}}(\eta) 
+ b_{\epsilon_{\rm em}}(\eta_\ast)P^{\{0\}}_{\epsilon_{g}\epsilon_{\rm em}}(\eta,\eta_\ast) \\ 
&\;\;\;\; + f_{\rm ion}(\eta,k^2)P^{\{0\}}_{\epsilon_{q}\epsilon_{\rm em}}(\eta,\eta_\ast) 
+ b_{\epsilon_{\rm em}}(\eta_\ast)f_{\rm ion}(\eta,k^2)P^{\{0\}}_{\epsilon_{\rm em}}(\eta_\ast)\,\,,
\end{split}
\end{equation}
while $P_{qq}$ is simply
\begin{equation}
\label{eq:P_qq}
P_{qq}(\eta,k) = b^2_{q}(\eta)P_{\rm L}(\eta,k) + P^{\{0\}}_{\epsilon_{q}}(\eta) 
+ 2 b_{\epsilon_{\rm em}}(\eta_\ast)P^{\{0\}}_{\epsilon_{q}\epsilon_{\rm em}}(\eta,\eta_\ast) 
+ b^2_{\epsilon_{\rm em}}(\eta_\ast)P^{\{0\}}_{\epsilon_{\rm em}}(\eta_\ast)\,\,.
\end{equation}
We notice that even if we stop at zeroth order in ${\cal H}\lambda_{\rm eff}$ we cannot constrain the parameter $b_{\epsilon_{\rm em}}$: 
a change in $b_{\epsilon_{\rm em}}$ can always be compensated by a change in the amplitude of $\epsilon_q$ 
in both \eq{P_gq} and \eq{P_qq}. This can be easily seen also at the level of the fields from \eq{q_tracers-B}. 

The importance of observations of the $q$ tracers lies in the (well-known: see \cite{Seljak:2008xr}) fact that, 
if we combine $P_{gg}$, $P_{gq}$ and $P_{qq}$, it is possible to beat down cosmic variance 
in the constraints on the different free bias parameters that $f_{\rm ion}$ depends on. In 
this case we expect this multi-tracer technique to be useful also because 
we are able to predict exactly the scale dependence of the different functions of $k$ that make up $f_{\rm ion}$, \ie~\eq{functions_f_n}: 
we are not expanding it in a power series in $k^2\lambda^2_{\rm eff}$, like we do with higher-derivative terms 
coming from gravitational dynamics (which are expanded in powers of $k^2 R^2(M_h)$). 

Still, it is also important to keep in mind that the situation is very different from that of constraints on local primordial non-Gaussianity, 
in the context of which the multi-tracer technique was originally proposed. 
Indeed, there the equivalent of $f_{\rm ion}$ is a function that scales as $k^{-2}$ at small $k$, and is therefore 
completely orthogonal to the higher-derivative corrections from gravitational evolution. A 
precise analysis on the usefulness of the multi-tracer technique is clearly beyond the scope of this paper, 
so we will not investigate this topic further here.

\subsection{More physics that can be captured in the bias expansion}
\label{ssec:caveats}

\noindent We can now check all the possible effects that we have neglected in the above discussion 
and see if and how they can affect the scale dependence of $f_{\rm ion}(\eta,k^2)$.

\subsubsection*{Integration over the past light cone}

\noindent The first and simplest one is the restriction of the spatial integral to the past light cone, $\abs{\vec{y}}\leq\eta-\eta_\ast$, in \eq{J_ion_approximated}. 
Including this restriction would modify the integral in \eq{redefine_f_ion-A} for every $n$. 
More precisely, it is straightforward to check that at a given $n$ that Fourier transform is now proportional to a 
dimensionless function of $k/(\eta-\eta_\ast)$, $k{\lambda}_{\rm eff}$ and $(\eta-\eta_\ast)/{\lambda}_{\rm eff}$, 
with a proportionality factor that still scales as ${\lambda}_{\rm eff}^{n\,+\,1}$. Therefore, 
the same reasoning applies: these functions can be computed 
without the need of any perturbative expansion in these three dimensionless variables, 
and increasing orders in $n$ are still suppressed by $({\cal H}{\lambda}_{\rm eff})^{n}$.

\subsubsection*{General expression for the optical depth}

\noindent Let us go back to \eqsII{hat_tau_approx}{integral_over_energy}: first, $\hat{\tau}(\eta,\abs{\vec{y}},E)$ takes the form 
\begin{equation}
\label{eq:time_dependence-A}
\begin{split}
\hat{\tau}(\eta,\abs{\vec{y}},E) &= \abs{\vec{y}}\int_{0}^{1}\dif u\,(n_\HI a)(\eta_\ast+u\abs{\vec{y}})\,
\sigma_{\rm bf}\big(E(\eta_\ast+u\abs{\vec{y}},\eta_\ast+\abs{\vec{y}})\big) \\
&= \underbrace{\abs{\vec{y}}\sigma_\infty\int_0^1\dif u\,(n_\HI a)(\eta_\ast+u\abs{\vec{y}})
\bigg(\frac{1+z(\eta_\ast+\abs{\vec{y}})}{1+z(\eta_\ast+u\abs{\vec{y}})}\bigg)^{\frac{7}{2}}}_{
\hphantom{\hat{\tau}_{\rm eff}(\eta,\,\abs{\vec{y}})\,}\equiv\,\hat{\tau}_{\rm eff}(\eta,\,\abs{\vec{y}})}\bigg(\frac{E_\infty}{E}\bigg)^{\frac{7}{2}}\,\,.
\end{split}
\end{equation}
With this definition, we can look in more detail to the integral over energy in the expression for $\delta n_g|_{\rm ion}$. 
As we discussed above \eq{integral_over_energy}, this integral can be carried out analytically (it can be written in terms of the incomplete Gamma function): 
while we could have used this function directly in Section \ref{ssec:single_flash}, we have approximated its behavior as that of an exponential. 
Now, using \eq{time_dependence-A} instead of \eq{hat_tau_approx}, we can write \eq{integral_over_energy} as 
\begin{equation}
\label{eq:time_dependence-B}
\int_{E_\infty}^{+\infty}\dif E\,\bigg(\frac{E}{E_\infty}\bigg)^{s\,-\,\frac{7}{2}}\,{e^{-\hat{\tau}(\eta,\,\abs{\vec{y}},\,E)}} \approx 
\frac{2E_{\infty}}{5-2s}\,e^{{-\frac{2s\,-\,5}{2(s\,-\,6)}}\hat{\tau}_{\rm eff}(\eta,\,\abs{\vec{y}})}\,\,,
\end{equation}
which again is correct for small $\hat{\tau}_{\rm eff}$ up to ${\cal O}(\hat{\tau}^2_{\rm eff})$ (see discussion below and Appendix \ref{app:appendix_energy_integral}). 
Let us then first see how we can treat the corrections that come from $\hat{\tau}_{\rm eff}(\eta,\,\abs{\vec{y}})$ not being simply proportional to $\abs{\vec{y}}$. 
These come from the dependence of the photon energy and the average hydrogen density on redshift, \ie~from the time dependence of the mean free path.

\subsubsection*{Time dependence of the mean free path}

\noindent It is relatively easy to deal with the time dependence of $n_\HI a$ and $E$ 
(which is reflected by the fact that $\hat{\tau}_{\rm eff}(\eta,\abs{\vec{y}})$ is an integral over the variable $u$). 
We can write the exponential term in \eq{time_dependence-B} as 
\begin{equation}
\label{eq:time_dependence-C}
\begin{split}
e^{{-\frac{2s\,-\,5}{2(s\,-\,6)}}\hat{\tau}_{\rm eff}(\eta,\,\abs{\vec{y}})} = e^{-\frac{\abs{\vec{y}}}{{\lambda}_{\rm eff}}}
\bigg[1&+\frac{\abs{\vec{y}}^2}{{\lambda}_{\rm eff}}\,\mathcal{O}\bigg(\frac{\partial\log{\lambda}_{\rm eff}(\eta_\ast)}{\partial\eta_\ast}\bigg) 
+\frac{\abs{\vec{y}}^2}{{\lambda}_{\rm eff}}\,\mathcal{O}\bigg(\frac{\partial\log(1+z_\ast)}{\partial\eta_\ast}\bigg)\bigg]\,\,.
\end{split}
\end{equation}
From this we see how these corrections give rise to the same type of terms that we already discussed 
in Section \ref{ssec:single_flash} (see \eg~\eq{delta_perturbations-B}). Those involving the derivative of $1+z_\ast$ scale as 
${\cal H}(\eta_\ast){\lambda}_{\rm eff} \sim {\cal H}{\lambda}_{\rm eff}$. 
Therefore, it is clear that at any fixed order in the expansion of $f_{\rm ion}(\eta,k^2)$ 
in ${\cal H}{\lambda}_{\rm eff}$ we only need a finite number of these terms: 
for example, the first of them modifies the scale dependence of the second term on the right-hand side of 
\eq{redefine_f_ion-B}. However, it is important to emphasize that these new terms do not involve any new free functions of time, 
so they affect $f_{\rm ion}(\eta,k^2)$ in a way which is completely under control. 

The same reasoning can, in principle, be applied to those involving the derivative of ${\lambda}_{\rm eff}$ at $\eta_\ast$. 
However, now we have 
\begin{equation}
\label{eq:time_dependence-D}
\begin{split}
\text{new terms from \eq{time_dependence-C}}
&\sim\bigg({\cal H}(\eta_\ast){\lambda}_{\rm eff}\frac{\partial\log{\lambda}_{\rm eff}(\eta_\ast)}{\partial\log\eta_\ast}\bigg)^n 
\sim\bigg({\cal H}{\lambda}_{\rm eff}\frac{\partial\log{\lambda}_{\rm eff}(\eta_\ast)}{\partial\log\eta_\ast}\bigg)^n\,\,.
\end{split}
\end{equation}
It is then clear that we can safely reabsorb these terms in the expansion of \eq{redefine_f_ion-B} 
only if ${\partial\log{\lambda}_{\rm eff}(\eta_\ast)}/{\partial\log\eta_\ast}$ is not much larger than $1$. 
This is not guaranteed to be true, since the mean free path can change rapidly during reionization. For example, if we assume 
that the neutral fraction evolves in a step-like fashion with a width $\delta\eta$, for $\eta_\ast$ close to the time at which reionization is halfway through we have that 
${\partial\log{\lambda}_{\rm eff}(\eta_\ast)}/{\partial\log\eta_\ast}\sim1/({\delta\eta\cal H}(\eta_\ast))$, 
which could be large if the time for the transition is much shorter than an Hubble time. 
For example, while Str{\" o}mgren spheres of ionized gas are forming, 
the m.f.p.~is fixed by the mean size of the bubbles, which can be very short until the bubbles 
coalesce. After this happens, and the universe reionizes, it is controlled by the mean density of the residual neutral hydrogen. 
In this scenario the size of the time derivatives of the m.f.p.~is 
controlled by how fast the coalescence happens with respect to a Hubble time.

Nevertheless we emphasize that, as before, the new terms of \eq{time_dependence-D} do not add new free functions 
of time to the expansion of $f_{\rm ion}(\eta,k^2)$. Even if the time evolution of the mean density of neutral hydrogen is still uncertain, this is 
different from our uncertainty on the microphysics of galaxy formation that is encoded by the Green's functions 
(\ie~by the bias coefficients $g_{n}$). Moreover, we expect that modeling the time dependence of the average hydrogen density $\widebar{n}_{\rm HI}(\eta)$ 
is much less difficult than that of, say, the fluctuations $\delta{n}_{\rm HI}(\eta,\vec{x})$ on the past light cone of the galaxies. 
In other words, we can capture these corrections non-perturbatively in ${\cal H}{\lambda}_{\rm eff}$ once we assume a time evolution of $\widebar{n}_\HI$.

\subsubsection*{Corrections from the integral over energy}

\noindent We then have to consider our approximation of the integral over energy in \eq{time_dependence-B}. 
We can account for the corrections to the right-hand side of that equation by using the full expression of the integral in terms of the incomplete Gamma function. 
This is shown in Appendix \ref{app:appendix_energy_integral}. 
The integral over $E$ shows the same qualitative behavior at large $\abs{\vec{y}}$, \ie~it goes to zero for $\abs{\vec{y}}\gg{\lambda}_{\rm eff}$. 
Therefore, nothing stops us from doing the same expansion of the Green's function in powers of $\abs{\vec{y}}$ that led to our expression for 
$f_{\rm ion}(\eta,k^2)$ as a power series in ${\cal H}{\lambda}_{\rm eff}$. The only difference will now be in the shape of the functions dimensionless 
functions of $k^2\lambda_{\rm eff}^2$ of \eqsII{redefine_f_ion-A}{functions_f_n}. We show these functions in \fig{functions_of_k_energy_integral} 
of Appendix \ref{app:appendix_energy_integral}. It is clear that their scale dependence is similar to those of \eq{functions_f_n}: 
they both go to $1$ at small $k$ and vanish at large $k$, with the turnaround being at $k\sim1/\lambda_{\rm eff}$.

\subsubsection*{Energy dependence of $\sigma_{\rm bf}$, $\varepsilon_{\rm em}$ and the Green's functions, and multiple absorbers}

\noindent The observation of the previous subsection is also what allows us to 
include the corrections to the bound-free cross section that we neglected in \eq{photoionization_cross_section_leading_scaling}. 
They do not change the behavior of the integral over energy. The qualitative dependence of the cross section on energy is still the same, 
\ie~it goes to zero for $E\gg E_\infty$ and scales as $(E_\infty/E)^{7/2}$ for small $E$. 
Therefore, if we neglect the time dependencies (that can be treated perturbatively anyway, as discussed above), 
the final result is still given by $E_\infty$ times a function of $\abs{\vec{y}}\sigma_\infty(n_\HI a)(\eta_\ast)$, 
where this function still goes to zero at very large $\abs{\vec{y}}$. The same conclusions then apply. 

We stress once again that in order to compute all the effects that we discussed here we require basically only the knowledge of the absorption cross section 
(that can be straightforwardly computed from first principles once we know what the absorbers are). 
Our uncertainty on the small-scale physics is only contained in the free functions $g_{n}$, 
exactly in the same way as the gravity-only bias expansion. 

Before proceeding to the next section, we briefly investigate what happens if we consider 
different species of absorbers, and different assumptions for the energy dependence of the emissivity or of the galaxy response: 
\begin{itemize}
\item including different absorbers with density $n_{\rm ab}$, each with their own 
absorption cross section $\sigma_{\rm ab}$, clearly affects the m.f.p., 
since now the optical depth $\tau$ is given by a sum over $\sigma_{\rm ab}n_{\rm ab}$ of terms like the one in \eq{optical_depth}. 
Therefore, as long as we know the energy dependence of the different $\sigma_{\rm ab}$, 
we can deal with this in the same way as we did when we considered only absorption from neutral hydrogen; 
\item things are a bit more complicated if we modify the energy dependence of the emissivity $\varepsilon_{\rm em}$ or the Green's functions in \eq{definition_of_G_g-B}. 
Let us assume, for example, that galaxy formation is sensitive to radiation at frequencies lower than $E_{\infty}$. 
Then, forgetting for a moment about redshift,\footnote{Depending on the difference in redshift between emission and absorption 
it is possible that some photons reach the galaxy with energies lower than the ionization energy, even if they were emitted with energies higher than $E_{\infty}$. 
The optical depth for these photons is then still controlled by the bound-free cross section for part of their path to the galaxy.} 
the optical depth for photons of such frequencies is controlled by the bound-bound cross section (\ie, mainly by the Rayleigh cross section). 
In this case, their effect on the bias expansion can still be treated in a similar way as we did above. 
More precisely, we can split the integral over $E$ in \eqsII{definition_of_G_g-A-1}{definition_of_G_g-A-2} 
in such a way that in each interval only one absorption line peaks. Then, for each of these sub-integrals it is possible to 
repeat the procedure above, ending up with a sum of different functions $f_{\rm ion}$: 
each of these functions will have its own specific m.f.p., which is controlled by the cross section at the corresponding absorption line; 
\item more problematic is instead the case where galaxy formation is sensitive to photons to which the universe is basically transparent, like for example X-rays, 
and the emissivity is significant at those frequencies (which is expected for active galactic nuclei and microquasars). 
In this case the effective m.f.p.~is of order of the Hubble radius, and we expect that any 
derivative expansion breaks down. A perturbative bias expansion would clearly be insufficient to describe 
this situation, so we will not investigate it further in this paper. 
\end{itemize}

\section{Full radiative transfer}
\label{sec:full_RT}

\noindent In this Section we see what happens if we drop basically all the assumptions on radiative transfer that we made in Section \ref{sec:radiative_bias}. We 
start by the considering a long epoch of emission of radiation on time scales of order a Hubble time and an inhomogeneous optical depth (Section \ref{ssec:generalization}). 
We show that now we need to add three functions of $k^2$ in the bias expansion (instead of one) 
in addition to the linear LIMD bias + stochastic term relation coming from gravitational interactions. 
The jump from one to two functions is due to the fact that now we need to take into account the dependence of the m.f.p.~and of the coefficients $g_{n}$ on $\eta_\ast$ 
(the latter comes just from the fact that we are expanding the Green’s function of galaxies around the emission time). 
Combined with the fact that the source overdensity is not evolving via a single growth factor because of the presence of the stochastic term $\epsilon_{\rm em}$, 
which evolves independently from the LIMD term $b_{\rm em}\delta$, 
we end up with the relation $\delta_g|_{\rm ion} = f_{\rm em}(k^2)\delta + f_{\epsilon_{\rm em}}(k^2)\epsilon_{\rm em}$. 
In the same way, adding inhomogeneities in the density of absorbers further increases the number of functions from two to three. 
This is essentially due to the fact that they play the role of sinks of radiation, \ie~of sources with negative emissivity. 
The most important point, however, is not the increase in the number of functions, but the fact 
that now it is impossible to predict the shape of these higher-derivative corrections to the bias expansion 
at all orders in ${\cal H}\lambda_{\rm eff}$, unlike what we found in Section \ref{ssec:single_flash}. 
It is enough to consider the case of homogeneous optical depth to see this. 
Indeed, now we need new response parameters for every significant fraction of a Hubble time during which the emissivity is not vanishing. 
Finally, in Section \ref{ssec:scatterings} we briefly study what is the impact of scattering on our analysis.

\subsection{Adding multiple flashes and inhomogeneities in the optical depth}
\label{ssec:generalization}

\noindent It is not difficult to see what happens if we move from the case of a very fast flash of radiation around $\eta = \eta_\ast$ 
to that of radiation emitted during an interval that can be of order of a Hubble time. 
In the case of a single emission, we have seen that at any given order $n$ in the ratio between the m.f.p.~and the Hubble radius 
the scale dependence of $f_{\rm ion}$ can be predicted at the price of $n$ 
free functions of time. Now, even if we fix $n$, as we increase the number of emission times we need more and more free coefficients 
$g_{n}$. In the limit of continuous emission that extends on cosmological time scales, we would then need an infinite number of bias parameters. 
In a sense (to be made more precise below), we are effectively integrating over $\eta_\ast$. 
The function $f_{\rm ion}$ depends on $\eta_\ast$ both through the time dependence of the m.f.p.~and, most importantly, 
through the coefficients $g_{n}$: since we do not know how they depend on time, we cannot carry out this integral. 

Let us discuss this in more detail. In the interest of clarity, we focus separately on inhomogeneities in the density of emitters and inhomogeneities in the optical depth.

\subsubsection*{Inhomogeneous emitting medium} 

\noindent We first consider two short bursts of radiation at $\eta_{\ast,1}$ and $\eta_{\ast,2}$, 
with $\eta_{\ast,1}<\eta_{\ast,2}$ and $\eta_{\ast,2}-\eta_{\ast,1}\gtrsim{\cal H}^{-1}$. 
Following the same steps that brought to \eq{J_ion_more_approximated-bis}, we now have 
(see also Appendix \ref{app:solution_single_flash} for details) 
\begin{equation}
\label{eq:sssec_generalization_source-A}
\begin{split}
\delta n_g|_{\rm ion}(\eta,\vec{x}) &= {\cal C}_{\rm ion}(\eta_{\ast,1})\widebar{\rho}_{\rm em}(\eta_{\ast,1}) 
\sum_{n\,=\,0}^{+\infty}g_{n}(\eta,\eta_{\ast,1}){\cal H}^{n} 
\int\dif^3y\,\abs{\vec{y}}^{n\,-\,2}\,\delta_{\rm em}(\eta_{\ast,1},\vec{x}+\vec{y})\,{e^{-\frac{\abs{\vec{y}}}{{\lambda}_{\rm eff}(\eta_{\ast,1})}}} \\
&\;\;\;\; + (\eta_{\ast,1}\to\eta_{\ast,2})\,\,,
\end{split}
\end{equation}
where we made explicit the dependence of ${\cal C}_{\rm ion}$ and $g_{n}$ on the emission time (we 
always intend $\cal H$ as evaluated at $\eta$, as in Section \ref{ssec:single_flash}). 
From the above equation we see that, in Fourier space, $\delta_g|_{\rm ion}$ is given by 
\begin{equation}
\label{eq:sssec_generalization_source-B}
\begin{split}
\delta_g|_{\rm ion}(\eta,\vec{k}) &= 
\frac{\widebar{n}_g|_{\rm ion}(\eta,\eta_{\ast,1})}{\widebar{n}_g|_{\rm ion}(\eta)}f_{\rm ion}(\eta,\eta_{\ast,1},k^2) 
\delta_{\rm em}(\eta_{\ast,1}\vec{k}) + (\eta_{\ast,1}\to\eta_{\ast,2})\,\,,
\end{split}
\end{equation}
where $f_{\rm ion}(\eta,\eta_{\ast,i},k^2)$ takes the same form as in \eq{delta_perturbations-B}, and $\widebar{n}_g|_{\rm ion}(\eta,\eta_{\ast,i})$ 
is the background galaxy number density computed for the single flash of radiation at $\eta_{\ast,i}$, 
so that $\widebar{n}_g|_{\rm ion}(\eta)$ is equal to $\widebar{n}_g|_{\rm ion}(\eta,\eta_{\ast,1})+\widebar{n}_g|_{\rm ion}(\eta,\eta_{\ast,2})$. 

Then, from \eq{sssec_generalization_source-B} we see that, in full generality, we can write $\delta_g|_{\rm ion}$ as the sum 
\begin{equation}
\label{eq:sssec_generalization_source-D}
\delta_g|_{\rm ion}(\eta,\vec{k}) = f_{\rm em}(\eta,k^2)\delta(\eta,\vec{k}) + f_{\epsilon_{\rm em}}(\eta,k^2)\epsilon_{\rm em}(\eta,\vec{k})\,\,.
\end{equation}
The two functions $f_{\rm em}$ and $f_{\epsilon_{\rm em}}$ are surely different since the matter 
overdensity and the stochastic term do not evolve with time in the same way. 
However, it is more interesting to ask if the ratio between them is scale-independent 
(as was the case when we considered a single emission time in Section \ref{ssec:galaxy_statistics}: see \eq{terms_in_bias_expansion}). 
The answer is negative. More precisely, since the time evolution of both $\delta$ and $\epsilon_{\rm em}$ is independent of $k$ 
at the order we are working at, this is due only to the dependence of the function $f_{\rm ion}$ on the emission time 
(through, \eg, the mean free path). 

Let us then see what happens in the limit of a continuous emission of radiation over an interval of order ${\cal H}^{-1}$. The 
question now is whether or not we can predict the scale dependence of the two functions multiplying $\delta$ and $\epsilon_{\rm em}$ in \eq{sssec_generalization_source-D}. 
The simplest way to see that the answer is negative is to focus on the leading order in the expansion in the mean free path. 
These two functions, then, depend on all the bias coefficients $g_{0}(\eta,\eta_{\ast,i})$ for $i=1,\dots,n_{\rm flashes}$ 
(generalizing the result of \eq{redefine_f_ion-B} to $n_{\rm flashes}$ emission times). 
As we bring $n_{\rm flashes}$ to infinity, we would then need an infinite number of functions of time, losing all predictivity. 

Before proceeding to study the inhomogeneities in the number density of absorbers, 
we notice that the solution for the intensity of radiation received by the galaxy 
in the limit of continuous emission over an interval $\eta_f-\eta_i\sim{\cal H}^{-1}$ can be obtained simply by an integral over $\eta_\ast$. 
Looking at the full solution of the RT equation in Appendix \ref{app:solution_of_boltzmann_equation}, 
it is straightforward to see that such limit corresponds to the replacement $\sum_{\eta_{\ast,i}}\Delta\eta\to\int_0^\eta\dif\eta_\ast$, 
where $\Delta\eta$ (defined at the beginning of Section \ref{ssec:single_flash}) is the duration of a single flash. 
However, we cannot obtain the dimensionless perturbation $\delta_g|_{\rm ion}$ in terms of an integral over $\eta_\ast$ as easily. 
The simplest way to see this is that $\Delta\eta$ always cancels in the expressions for $\delta_g|_{\rm ion}$ (as shown in \eq{redefine_f_ion-B}, for example).

\subsubsection*{Inhomogeneous absorbing medium}

\noindent We know that there are inhomogeneities in the number density of neutral hydrogen. 
The ionization fronts propagating into the neutral medium formed Str{\" o}mgren spheres of ionized, heated gas, 
so we cannot treat $n_\HI$ as homogeneous on scales of order of the typical bubble size, 
but we have to take into account $\delta_\HI$ for scales $k\gtrsim\kmpc{d-2}$ \cite{Mao:2014rja}. 
We can imagine that these inhomogeneities in the absorbing medium can also be treated by a bias expansion on sufficiently large scales. 
Indeed, \cite{McQuinn:2018zwa} has recently developed such a bias expansion for the inhomogeneities in the neutral fraction 
and, with it, a bias expansion for the $\rm 21cm$ signal from reionization (see also \cite{Giri:2018dln} for a recent 
computation of the position-dependent $\rm 21cm$ power spectrum via separate-universe simulations). 

Since we have the full solution of the RT equation (\eqsIII{boltzmann_solution}{source_definition}{tau_solution} in Appendix \ref{app:solution_of_boltzmann_equation}), 
it should be easy, in principle, to study the effect of these inhomogeneities in the optical depth. 
However, since $\tau$ itself is given by an integral along the line of sight, 
the resulting expressions for $\widebar{n}_g|_{\rm ion}$ and $\delta n_g|_{\rm ion}$ are very complicated. 
For this reason, we study the effect of an inhomogeneous $\tau$ in the case of radiation emitted in a single flash around $\eta_\ast$: 
this is simpler to treat, and the generalization to continuous emission is straightforward. 

Let us go back to \eqsII{observed_intensity}{optical_depth}. We can expand $n_\HI$ in a background $\widebar{n}_\HI$ and a dimensionless perturbation $\delta_\HI$. 
We see that, at first order in perturbations, the contribution of the $\tau$ inhomogeneities to the perturbations $\delta{\cal I}$ 
of the specific intensity of ionizing radiation received by the galaxies is equal to 
\begin{equation}
\label{eq:sssec_generalization_tau}
\delta{\cal I}\supset{-\bigg(\frac{1+z(\eta)}{1+z_\ast}\bigg)^3} e^{-\widebar{\tau}}\,\widebar{{\cal I}}_\ast\big(\eta_\ast,E(\eta_\ast,\eta)\big) 
{\int_{\eta_\ast}^\eta}\dif\eta'\,\frac{\delta_\HI\big(\eta',\vec{x}+\vers{n}(\eta-\eta')\big)}{\lambda_{\rm ion}\big(\eta',E(\eta',\eta)\big)}\,\,,
\end{equation}
where the energy-dependent m.f.p.~is given, as in \eq{hat_tau_approx}, by $\lambda_{\rm ion} = 1/(\sigma_{\rm bf}\widebar{n}_\HI a)$, 
and $\widebar{\tau}$ is obtained from \eq{optical_depth} by taking $n_\HI=\widebar{n}_\HI$.~Then, 
we see that the structure of \eq{sssec_generalization_tau} is similar to that of the full solution of Appendix \ref{app:solution_of_boltzmann_equation}, 
if we take a homogeneous optical depth and an emissivity proportional to $\sigma_{\rm bf} \widebar{n}_\HI$. 
That is, the absorber medium plays a role of an additional ``negative'' source term (\ie~a sink), 
whose emissivity is not localized around a single time $\eta_\ast$ (unlike that of the emitters). 

Generalizing to an arbitrary number of emission times, this tells us that the inhomogeneities in the density of absorbers contribute to $\delta_g|_{\rm ion}$ 
through two additional functions of $(\eta,k^2)$, which we can call $f_\HI$ and $\fHI$. These two functions multiply $\delta$ and $\epsilon_{\rm HI}$ respectively. 
There is no loss of generality in summing $f_\HI$ and $f_{\rm em}$ into a single function, given that both multiply the matter overdensity $\delta$. 
We cannot instead reabsorb $\fHI$ in $f_{\epsilon_{\rm em}}$ since the two stochastic terms $\epsilon_\HI$ and $\epsilon_{\rm em}$ are different fields. 
However, the most important point to emphasize is that, again, the scale dependence of $f_\HI$ and $\fHI$ 
cannot be computed perturbatively with a finite number of bias parameters.

\subsection{Scattering} 
\label{ssec:scatterings} 

\noindent While a full discussion of the impact of scattering on the radiative-transfer equation is beyond the scope of this paper, 
in this short section we briefly draw a qualitative picture of how our previous results would be affected if they play an important role. 

The most important effect is the redistribution of the direction of the photons.\footnote{The scattering kernel 
can be only a function of $\vers{n}\cdot\vers{n}'$ at the order in perturbations we are working at. 
The reason is the same as the one we used to conclude that the emissivity and the galaxy response cannot depend on $\vers{n}$.} 
Let us then consider again the case of a single flash to build some intuition. 
We can imagine that, after photons are emitted isotropically at $\eta_\ast$, some of those that are initially directed 
away from the galaxy are later scattered towards it at some time $\eta_{\rm sc} > \eta_\ast$. 
This effectively mimics a second burst of radiation at $\eta_{\rm sc}$. Therefore, in the limit of scattering happening continuously over time, 
we expect to roughly go back to the same situation that we have discussed when we have included inhomogeneities in the optical depth. 

Finally, what happens if scattering dominates over absorption? 
In this case we expect that the intensity is made isotropic on scales larger than the (comoving) diffusion length $D\sim\sqrt{1/(\sigma_{\rm sc}\widebar{n}_{\rm sc}a\cal H)}$ 
($\sigma_{\rm sc}$ and $\widebar{n}_{\rm sc}$ being the overall amplitude of the scattering cross section and the average number density of scatterers, respectively). 
On these scales, the monopole of the intensity will follow a (sourced) diffusion equation, so once we compute $\delta_g|_{\rm ion}$ 
using \eqsII{definition_of_G_g-A-1}{definition_of_G_g-A-2} we see that the higher-derivative terms from RT effects can be expanded in powers of $k^2D^2$.

\section{Discussion and conclusions}
\label{sec:conclusions}

\noindent In this paper we have investigated whether it is possible to find a resummation of the higher-derivative terms in the bias expansion that come from 
radiative-transfer (RT) effects. 

We have shown that, at linear order in perturbations and when mainly absorption and emission play a relevant role in the RT equation, these 
effects are captured by three functions of $k^2\lambda_{\rm eff}^2$, where $\lambda_{\rm eff}$ is the effective mean free path of ionizing radiation. 
Whether or not it is possible to predict the scale dependence of these three functions, 
instead of simply relying on their expansion in powers of $k^2\lambda_{\rm eff}^2$, depends on the following factors: 
\begin{itemize}
\item the time dependence of the emissivity of the sources of ionizing radiation; 
\item the time dependence of the response of galaxies to the flux of ionizing radiation; 
\item the presence of inhomogeneities in the optical depth. 
\end{itemize}

\begin{table}[b!]
\myfloatalign
\caption[.]{This table shows which assumptions are necessary to predict the scale dependence of the bias from RT effects, assuming no inhomogeneities in the optical depth. 
$\Delta\eta_{\rm em}$ and $\Delta\eta_{G}$ are the interval over which $\varepsilon_{\rm em}$ is non-vanishing and the typical extent in time of the galaxy response, respectively. 
By~\cmark~we mean that the higher-derivative terms can be resummed into well-defined functions of $k^2\lambda_{\rm eff}^2$, each multiplied by an RT bias coefficient 
and an increasing power of ${\cal H}\lambda_{\rm eff}$. The case of an instantaneous galaxy response is discussed in \mbox{Appendix \ref{app:fourier_transform_intensity}.}} 
\label{tab:conclusions_table-1}
\centering
\medskip
\begin{tabular}{lcc}
\toprule
{$\delta\tau=0$} & $\Delta\eta_{G}\ll {\cal H}^{-1}$ & $\Delta\eta_{G}\sim {\cal H}^{-1}$ \\
\midrule
$\Delta\eta_{\rm em}\ll {\cal H}^{-1}$ & \cmark & \cmark \\[2ex]
$\Delta\eta_{\rm em}\sim {\cal H}^{-1}$ & \cmark & \xmark \\
\bottomrule
\end{tabular}
\end{table}

\begin{table}[b!]
\myfloatalign
\caption[.]{Same as \tab{conclusions_table-1}, but taking into account the inhomogeneities in the optical depth. 
Note that having an inhomogeneous $\tau$ is not a problem if the galaxy response is instantaneous: 
we refer to Appendix \ref{app:fourier_transform_intensity} for more details.}
\label{tab:conclusions_table-2}
\centering
\medskip
\begin{tabular}{lcc}
\toprule
{$\delta\tau\neq 0$} & $\Delta\eta_{G}\ll {\cal H}^{-1}$ & $\Delta\eta_{G}\sim {\cal H}^{-1}$ \\
\midrule
$\Delta\eta_{\rm em}\ll {\cal H}^{-1}$ & \cmark & \xmark \\[2ex]
$\Delta\eta_{\rm em}\sim {\cal H}^{-1}$ & \cmark & \xmark \\
\bottomrule
\end{tabular}
\end{table}

This is summarized in \tabs{conclusions_table-1}{conclusions_table-2}. For example, let us consider the case of the galaxy 
response varying on cosmological time scales; that is, the observable properties of a given galaxy sample retain a memory of the ionizing flux received at some earlier time. 
Then, the dependence of these functions on the details of galaxy formation can be absorbed in 
generalized RT bias coefficients, without needing a detailed ``UV'' modeling of galaxy formation, if: \emph{a}) one can neglect inhomogeneities in the optical depth; 
\emph{b}) the emission of ionizing radiation only happens for a short period of time $\Delta\eta\ll{\cal H}^{-1}$. 
If these assumptions do not hold, we cannot predict the dependence on $k$ of these three functions 
unless we know precisely how the response of galaxies to the ionizing radiation depends on time. 
It is important to stress that this does not happen in the gravity-only bias expansion. 
There, the time dependence of the Green's function can be reabsorbed, order by order in perturbations and spatial derivatives, 
into the time-dependent bias coefficients. Here we cannot do this because, through the received flux, 
the galaxies are sensitive to the inhomogeneities in the distribution of sources and sinks of ionizing radiation evaluated along 
their past light cone, and not only along the past fluid worldline. 
Consequently, the time dependence of the Green's function affects also the scale dependence of the bias. 

It is however important to emphasize that it is entirely possible for the response to ionizing radiation to happen on time scales much 
faster than a Hubble time. For example, in the case of line emission from diffuse gas we can imagine that the response to the ambient radiation field 
is controlled by the recombination rate, which has nothing to do with Hubble. 
More precisely, if we wanted to be completely general, we should have considered a Fourier transform of the Green's functions of Section \ref{ssec:single_flash} 
with respect to $\eta-\eta'$. This Fourier transform can have support for both $\omega\sim{\cal H}$ and for $\omega\gg{\cal H}$, 
and in this work we have focused on the case where the Fourier coefficients are nonzero at low frequencies. 
In this sense, our study is conservative: as we can see from \tabs{conclusions_table-1}{conclusions_table-2}, 
the contributions to the bias expansion due to the high-frequency part of the response are always under control. 

In this paper we have followed an effective field theory approach, \ie~we have not made any specific assumption on the 
Green's functions describing the response to the radiation (apart from very general assumptions on their time dependence, as discussed above). 
In other words, our results apply equally well to any other tracer of the underlying matter distribution that is sensitive to RT effects. 

For galaxies, the suppression of the star-formation rate coming from photo-evaporation of the gas accreting onto the parent halo 
is not very relevant for halos whose mass is much larger than the Jeans mass. 
We thus expect that the bias coefficients of the higher-derivative terms from these RT effects are actually a strong function of the parent halo mass, 
and their amplitude is very small for $M_{h}\gg M_{\rm J}$. Since we do expect the RT to typically lead to a smooth contribution to the 
galaxy power spectrum (see \fig{PS_corrections}), a contribution that is very small in amplitude can presumably 
be absorbed by marginalizing over a sufficiently flexible template for the RT effects. 
However, this might well lead to a degradation of constraints on the neutrino mass $m_\nu$ or equilateral non-Gaussianity $f_{\rm NL}^{\rm equil}$. 
In particular, \fig{neutrino_scale_dependence} illustrates that the RT effects are expected to have a scale dependence 
that is quite similar to that induced by nonzero neutrino masses. 

However, we do not expect RT effects to be necessarily small for tracers like, \eg, the Lyman-$\alpha$ forest. In this case, we can imagine that 
treating the higher-derivative terms perturbatively (\ie~through an expansion in powers of $k^2$) would to lead to a significant loss of 
constraining power on cosmological parameters in general. Indeed, the bias expansion would stop being predictive already at 
the large scales where the RT effects leave their imprint on the power spectrum of these tracers. 

\paragraph{Future prospects} An explicit modeling of the response to ionizing radiation is surely a way to go around this problem. 
However, it is clear that any uncertainty on the theoretical errors of this model would reduce 
the robustness of galaxy clustering as a cosmological probe. A second way relies on the fact that 
we do not expect RT effects to induce a sizeable velocity bias. Indeed, let us consider the momentum density $\vec{p}_{\cal I}$ carried 
by the ionizing radiation field, as seen by an observer comoving with the galaxies. 
It is straightforward to show that, at leading order in derivatives, it is proportional to $\widebar{\rho}_{\cal I}\lambda_{\rm eff}\vec{\nabla}\delta$ 
(see Appendix \ref{app:momentum}), where $\widebar{\rho}_{\cal I}$ is the average energy density of the radiation field. 
The momentum transfer from the radiation field to the galaxies, and consequently the contribution from RT to 
the velocity bias $\vec{v}_g-\vec{v}$, is then controlled by the ratio $\widebar{\rho}_{\cal I}/\widebar{\rho}_b = \Omega_{\cal I}/\Omega_b$. 
Since the density of radiation is dominated by the CMB, this ratio is much smaller than $\Omega_r/\Omega_b\approx 5\Omega_r/\Omega_m\approx 10^{-3}(1+z)$. 
From this we conclude that the peculiar velocities of galaxies, and with them the higher-multipole contribution 
to the galaxy overdensity in redshift space, are very weakly affected by radiative transfer 
(as pointed out by \cite{Gontcho:2014nsa}), allowing us to obtain, in principle, unbiased measurements on cosmological parameters. 

\paragraph{Comparison with recent work} While this work was being completed, 
two papers that investigate similar topics have been published on the arXiv \textnormal{\cite{Meiksin:2018wij,Sanderbeck:2018lwc}}. In 
\cite{Meiksin:2018wij} the authors compute the inhomogeneities $\delta_{\cal I}$ of the intensity $\cal I$ 
(integrated over angles and averaged over frequencies with some weighting $w(E)$, whose frequency dependence is 
assumed to be that of the $\HI$ photoionization cross section). 
The spirit of our calculation is similar to theirs (see for example Appendix \ref{app:fourier_transform_intensity}), with the difference that they 
consider directly the physically-motivated case of continued emission over Hubble time scales. 
The work \cite{Sanderbeck:2018lwc}, instead, studies the impact of ionizing radiation on the clustering of tracers. 
The difference with our computation is that the authors assume that the contribution of radiative transfer, 
\ie~what we called $\delta_g|_{\rm ion}$, is given by $b_{\cal I}\delta_{\cal I}$, with $b_{\cal I}$ 
a time-dependent bias coefficient (see their Eqs.~(1), (2)). That is, 
they assume that the scale of nonlocality in time of the response of tracers to ionizing radiation is much faster than Hubble, 
so that their Green's functions are proportional to $b_{\cal I}(\eta)\delta(\eta-\eta')$. 
Consequently, based on this strong simplifying assumption and after assuming a model for the emissivity and the absorption coefficient, 
they are able to compute the scale dependence of these corrections to galaxy clustering without requiring an infinite number of bias coefficients. 
This is in agreement with our conclusions (see \eg~\tabs{conclusions_table-1}{conclusions_table-2}).

\section*{Acknowledgements}

\noindent It is a pleasure to thank Philipp Busch, Chris Byrohl, Jens Chluba, Ildar Khabibullin, Eiichiro Komatsu, Kaloian Lozanov, Matt McQuinn, and Shun Saito for useful discussions. 
G.~C. and F.~S. acknowledge support from the Starting Grant (ERC-2015-STG 678652) ``GrInflaGal'' from the European Research Council.

\appendix

\section{Review of radiative-transfer physics}
\label{app:radiative_transfer}

\noindent In this appendix we review the solution of the radiative-transfer equation in an FLRW universe 
and give some details on the calculations of Sections \ref{sec:radiative_bias} and \ref{sec:conclusions}.

\subsection{Solution of Boltzmann equation}
\label{app:solution_of_boltzmann_equation}

\noindent In Section \ref{ssec:emission_absorption} we discussed in detail the emission and absorption coefficients, 
and derived the equation of radiative transfer, \ie~\eq{boltzmann-3} (with \eq{boltzmann-1} being its fully general-relativistic version). 
This equation can be straightforwardly solved via an integral along the line of sight. More precisely, the phase-space density can \mbox{be written as} 
\begin{equation}
\label{eq:boltzmann_solution}
\begin{split}
&{\cal I} = \int_{0}^\eta\dif\eta'\,\bigg(\frac{1+z(\eta)}{1+z(\eta')}\bigg)^3e^{{-\tau}}{\cal S}\big(\eta',\vec{x}+\vers{n}(\eta-\eta'),E(\eta',\eta)\big)\,\,,
\end{split}
\end{equation}
where we defined the source term $\cal S$ as 
\begin{equation}
\label{eq:source_definition}
{\cal S}=\frac{\rho_{\rm em}\varepsilon_{\rm em} a}{4\pi}\,\,.
\end{equation}
In \eq{boltzmann_solution} we integrate starting from $\eta'=0$ since the emissivity vanishes for times earlier than some $\eta_i$ in any case. 
The solution for the optical depth $\tau$, which enters in the integral of \eq{boltzmann_solution}, is 
\begin{equation}
\label{eq:tau_solution}
\tau = \int_{\eta'}^{\eta}\dif\eta''\,(\sigma_{\rm ab}n_{\rm ab} a)\big(\eta'',\vec{x}+\vers{n}(\eta-\eta''),E(\eta'',\eta)\big)\,\,.
\end{equation}
In both \eqsII{boltzmann_solution}{tau_solution}, the evolution of the energy is given by
\begin{equation}
\label{eq:energy_solution}
E(\eta',\eta)=E\frac{1+z(\eta')}{1+z(\eta)}\,\,.
\end{equation}

\subsection{Solution for single flash}
\label{app:solution_single_flash}

\noindent Let us now study in more detail the solution in the case of a single flash. 
The assumption that the emissivity is nonzero only for a short interval $\Delta\eta\ll{\cal H}^{-1}$ around $\eta_\ast$, 
which we obviously take to be less than $\eta$ in \eq{boltzmann_solution}, amounts to writing 
\begin{equation}
\label{eq:delta_dirac_emissivity}
\varepsilon_{\rm em}(\eta, E) = \Delta\eta\,\delta(\eta-\eta_\ast)\,\varepsilon_{{\rm em},\ast}(E)\,\,.
\end{equation}
Plugging this into \eq{boltzmann_solution} we arrive directly at \eq{observed_intensity}, where ${\cal I}_{\ast}$ is indeed given by 
\begin{equation}
\label{eq:source_to_initial_intensity}
{\cal I}_{\ast} = \frac{\Delta\eta\,\varepsilon_{{\rm em},\ast}(E)}{4\pi}(\rho_{\rm em}a)|_{\eta=\eta_\ast}\,\,,
\end{equation}
as we wrote below \eq{observed_intensity} (noticing that we called $\varepsilon_{{\rm em},\ast}(E) = \varepsilon_{{\rm em}}(\eta_\ast,E)$ in Section \ref{ssec:single_flash}). 
The generalization to multiple quick flashes of duration $(\Delta\eta)_i$ at different times $\eta_{\ast,i}$, 
which has been used in Section \ref{ssec:generalization}, is straightforward: 
one can just take the emissivity to be a sum of terms like that of \eq{delta_dirac_emissivity}. 
Moreover, it is also easy to see that we can recover the full solution of \eq{boltzmann_solution} 
by integrating the solution for a single flash in $\dif\eta_\ast$. 

We are now in the position to check in detail how to go from \eq{definition_of_G_g-A-2} to \eq{J_ion}. Using 
\eqsV{observed_intensity}{definition_of_G_g-B}{reabsorb_redshift}{G_g_expansion}{source_to_initial_intensity}, we see that $\delta n_g|_{\rm ion}$ is given by 
\begin{equation}
\label{eq:app_single_flash-A}
\begin{split}
\delta n_g|_{\rm ion}(\eta,\vec{x}) = \frac{\Delta\eta\,a_\ast\,{\cal G}}{4\pi}\int_{\eta_\ast}^\eta\dif\eta'\int\dif\vers{n}\int_0^{+\infty}\dif E&\,
{G}^{(1)}_{g}(\eta,\eta')\,\sigma_{\rm bf}(E)\,e^{-\tau}\,\varepsilon_{{\rm em},\ast}\big(E(\eta_\ast,\eta')\big)\,\times \\
&\,\delta\rho_{\rm em}\big(\eta_\ast,\vec{x}+\vers{n}(\eta'-\eta_\ast)\big)\,\,,
\end{split}
\end{equation}
where the integral over energy actually starts from $E_\infty$ and the optical depth $\tau$ at all orders in perturbations is given by (see \eq{optical_depth})
\begin{equation}
\label{eq:app_single_flash-B}
\tau = \int_{\eta_\ast}^{\eta'}\dif\eta''\,(\sigma_{\rm bf}n_\HI a)\big(\eta'',\vec{x}+\vers{n}(\eta'-\eta''),E(\eta'',\eta')\big)\,\,.
\end{equation}
To see that the integral in $\dif\eta'\dif\vers{n}$ is just an integral over $\dif^3y/\abs{\vec{y}}^2$ we redefine $\eta'-\eta_\ast\equiv y$ 
(where $y$ will be equal to $\abs{\vec{y}}$ at the end). 
Then, we have (dropping the overall proportionality factor $({\Delta\eta\,a_\ast\,{\cal G}})/({4\pi})$ to simplify the notation) 
\begin{equation}
\label{eq:app_single_flash-C}
\begin{split}
\delta n_g|_{\rm ion}(\eta,\vec{x})&\propto\int_{\eta_\ast}^\eta\dif\eta'\int\dif\vers{n}\int_0^{+\infty}\dif E\,
{G}^{(1)}_{g}(\eta,\eta')\,\sigma_{\rm bf}(E)\,\varepsilon_{{\rm em},\ast}\big(E(\eta_\ast,\eta')\big)\,\times \\
&\hphantom{\,\,\propto\int_{\eta_\ast}^\eta\dif\eta'\int\dif\vers{n}\int_0^{+\infty}\dif E\, } 
\delta\rho_{\rm em}\big(\eta_\ast,\vec{x}+\vers{n}(\eta'-\eta_\ast)\big)\,\times \\
&\hphantom{\,\,\propto\int_{\eta_\ast}^\eta\dif\eta'\int\dif\vers{n}\int_0^{+\infty}\dif E\, } 
e^{{-\int}^{\eta'}_{\eta_\ast}\dif\eta''\,(\sigma_{\rm bf}n_\HI a)(\eta'',\,\vec{x}\,+\,\vers{n}(\eta'\,-\,\eta''),\,E(\eta'',\,\eta'))} \\
&=\int_{0}^{\eta\,-\,\eta_\ast}\dif y\int\dif\vers{n}\int_0^{+\infty}\dif E\,
{G}^{(1)}_{g}(\eta,\eta_\ast+y)\,\sigma_{\rm bf}(E)\,\varepsilon_{{\rm em},\ast}\big(E(\eta_\ast,\eta_\ast+y)\big)\,\times \\
&\hphantom{\propto\int_{0}^{\eta\,-\,\eta_\ast}\dif\eta'\int\dif\vers{n}\int_0^{+\infty}\dif E\, } 
\delta\rho_{\rm em}(\eta_\ast,\vec{x}+\vers{n}y)\,\times \\
&\hphantom{\propto\int_{0}^{\eta\,-\,\eta_\ast}\dif\eta'\int\dif\vers{n}\int_0^{+\infty}\dif E\, } 
e^{{-\int}^{\eta_\ast\,+\,y}_{\eta_\ast}\dif\eta''\,(\sigma_{\rm bf}n_\HI a)(\eta'',\,\vec{x}\,+\,\vers{n}(\eta_\ast\,+\,y\,-\,\eta''),\,E(\eta'',\,\eta_\ast\,+\,y))}\,\,.
\end{split}
\end{equation}
Then, in the integral for $\tau$ we define $\eta''=\eta_\ast+uy$, so that \eq{app_single_flash-C} becomes
\begin{equation}
\label{eq:app_single_flash-D}
\begin{split}
\delta n_g|_{\rm ion}(\eta,\vec{x}) 
&\propto\int_{0}^{\eta\,-\,\eta_\ast}\dif y\int\dif\vers{n}\int_0^{+\infty}\dif E\,
{G}^{(1)}_{g}(\eta,\eta_\ast+y)\,\sigma_{\rm bf}(E)\,\varepsilon_{{\rm em},\ast}\big(E(\eta_\ast,\eta_\ast+y)\big)\,\times \\
&\hphantom{\propto\int_{0}^{\eta\,-\,\eta_\ast}\dif\eta'\int\dif\vers{n}\int_0^{+\infty}\dif E\, } 
\delta\rho_{\rm em}(\eta_\ast,\vec{x}+\vers{n}y)\,\times \\
&\hphantom{\propto\int_{0}^{\eta\,-\,\eta_\ast}\dif\eta'\int\dif\vers{n}\int_0^{+\infty}\dif E\, } 
e^{{-y}\int^{1}_{0}\dif u\,(\sigma_{\rm bf}n_\HI a)(\eta_\ast\,+\,uy,\,\vec{x}\,+\,\vers{n}(1\,-\,u)y,\,E(\eta_\ast\,+\,uy,\,\eta_\ast\,+\,y))}\,\,.
\end{split}
\end{equation}
Using $\vers{n}y=\vec{y}$, and $\dif y\dif\vers{n}=\dif^3y/\abs{\vec{y}}^2$, we finally recognize \eq{J_ion}. 
Moreover, we see that $\hat{\tau}$ is given by 
\begin{equation}
\label{eq:single_flash-E}
\begin{split}
\hat{\tau}(\eta,\vec{x},\vec{y},E) &= \abs{\vec{y}}\int_{0}^{1}\dif u\,(\sigma_{\rm bf}n_\HI a)\big(\eta_\ast+u\abs{\vec{y}},\vec{x}+(1-u)\vec{y}, 
E(\eta_\ast+u\abs{\vec{y}},\eta_\ast+\abs{\vec{y}})\big) \\
&= \abs{\vec{y}}\int_{0}^{1}\dif u\,(n_\HI a)\big(\eta_\ast+u\abs{\vec{y}},\vec{x}+(1-u)\vec{y}\big)\,
\sigma_{\rm bf}\big(E(\eta_\ast+u\abs{\vec{y}},\eta_\ast+\abs{\vec{y}})\big)\,\,,
\end{split}
\end{equation}
so that if we assume a homogeneous density of absorbers (\ie~of $1s$ hydrogen, in this case), we indeed find \eq{hat_tau}.

\subsection{Full integral over energy}
\label{app:appendix_energy_integral}

\begin{figure}
\centering
\includegraphics[width=0.85\columnwidth]{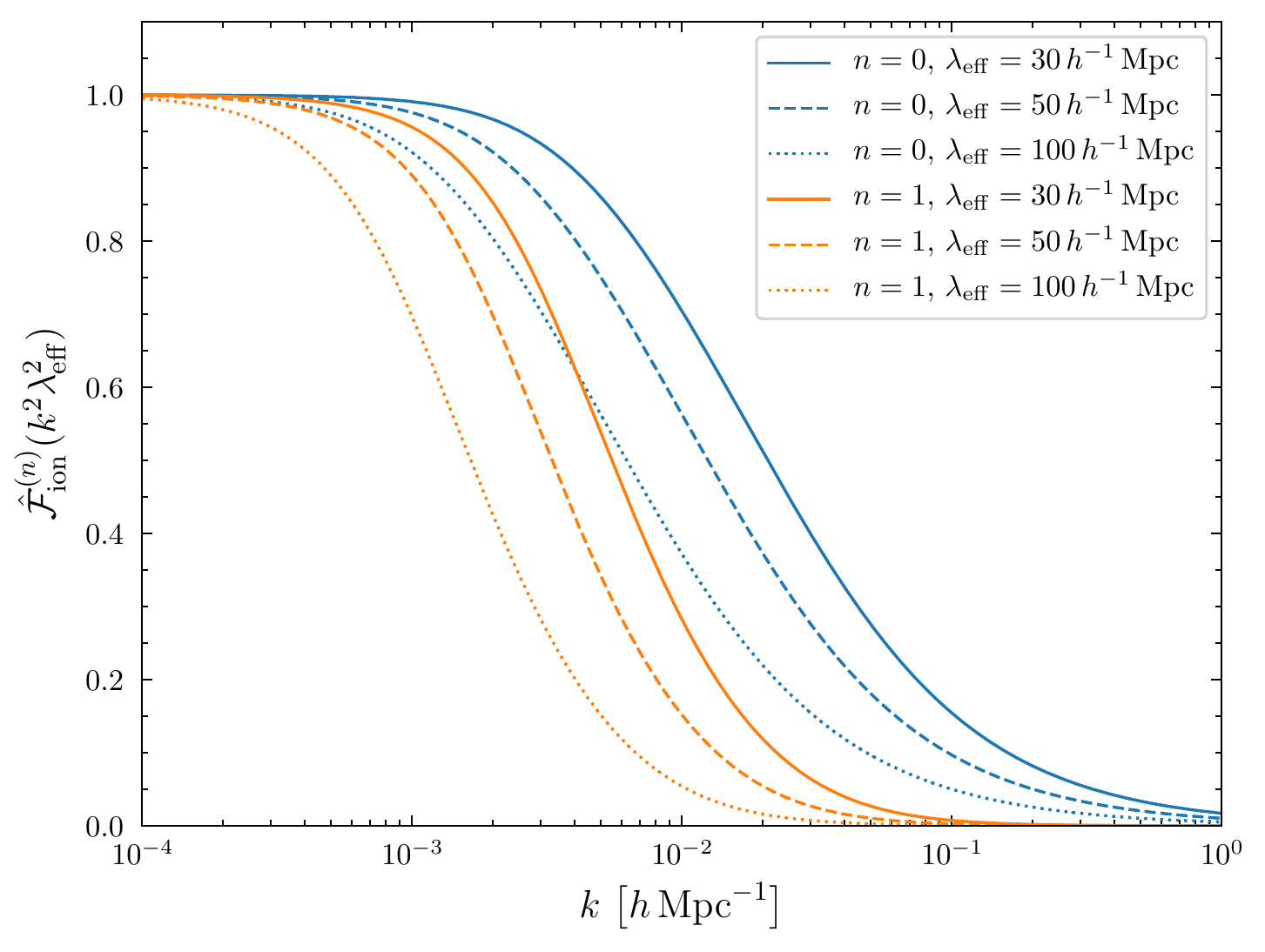} 
\captionsetup{singlelinecheck=off}
\caption[.]{Plot of the first two functions $\smash{\hat{{\cal F}}^{(n)}_{\rm ion}}$ that derive from \eq{appendix_energy-A} 
for different values of the effective m.f.p.~(we call them $\smash{\hat{{\cal F}}}^{(n)}_{\rm ion}$ to distinguish them from those of \eq{functions_f_n}). 
$\widetilde{E}/E_\infty$ is fixed to $2$. At large $k\gg 1/\lambda_{\rm eff}$ they decay as $1/k^{n\,+\,2/7}$.} 
\label{fig:functions_of_k_energy_integral}
\end{figure}

\noindent In this appendix we compute the full integral of \eq{time_dependence-B} and discuss the corresponding corrections to the 
scale dependence of the bias. In order to do this, we first make the (physically motivated) assumption that the emissivity vanishes above some energy $\widetilde{E}>E_\infty$. 
Further, we assume that its spectral index $s$ is equal to zero, for simplicity. Dropping all corrections coming from the redshift dependence of the photon energy 
and the time dependence of the mean free path (which are discussed in detail in Section \ref{ssec:caveats}), \eq{time_dependence-B} 
(or, equivalently, \eq{integral_over_energy}) becomes
\begin{equation}
\label{eq:appendix_energy-A}
\begin{split}
\int_{E_\infty}^{\widetilde{E}}\dif E\,\bigg(\frac{E}{E_\infty}\bigg)^{-\frac{7}{2}}\,{e^{-\frac{\abs{\vec{y}}}{\lambda_{\rm ion}(\eta_\ast,\,E)}}} &= 
\frac{2E_\infty}{7}\bigg(\frac{\abs{\vec{y}}}{\lambda_{\rm eff}(\eta_\ast)}\bigg)^{-\frac{5}{7}}\,\times \\
&\;\;\;\;\bigg[\Gamma\bigg(\frac{5}{7},\frac{\abs{\vec{y}}/\lambda_{\rm eff}(\eta_\ast)}
{(\widetilde{E}/E_\infty)^{7/2}}\bigg)-\Gamma\bigg(\frac{5}{7},\frac{\abs{\vec{y}}}{\lambda_{\rm eff}(\eta_\ast)}\bigg)\bigg]\,\,,
\end{split}
\end{equation}
where $\lambda_{\rm ion}(\eta_\ast,E)$ is defined by \eqsII{photoionization_cross_section_leading_scaling}{hat_tau_approx}, 
and in the rest of this appendix we define the effective m.f.p.~as $\lambda^{-1}_{\rm eff}(\eta_\ast)\equiv \sigma_\infty(n_\HI a)(\eta_\ast)$ 
(and drop its dependence on $\eta_\ast$ to avoid cluttering the notation). 

If we then follow \eq{J_ion_more_approximated} and define the dimensionless parameter ${\cal C}_{\rm ion}$ as 
\begin{equation}
\label{eq:appendix_energy-B}
{\cal C}_{\rm ion}\equiv \frac{\Delta\eta\,a_\ast\,{\cal G}\,\sigma_\infty\,\varepsilon_\infty\,E_\infty}{10\pi}\bigg[1-\bigg(\frac{E_\infty}{\widetilde{E}}\bigg)^{\frac{5}{2}}\bigg]\,\,,
\end{equation}
we can read off from \eq{appendix_energy-A} the equivalent of the exponential $\exp({-\abs{\vec{y}}/\lambda_{\rm eff}})$. 
This function has the same behavior of $\exp({-\abs{\vec{y}}/\lambda_{\rm eff}})$: it goes to $1$ for $\abs{\vec{y}}\ll\lambda_{\rm eff}$, 
and decays exponentially at large $\abs{\vec{y}}$.\footnote{In order to have this exponential decay it is necessary 
that the integral over energy does not go up to $E=+\infty$. This is why in \eq{appendix_energy-A} we have assumed that the emissivity vanishes for $E > \widetilde{E}$.} 
Therefore, following the same steps that lead to \eqsII{redefine_f_ion-A}{redefine_f_ion-B}, we can define the equivalent of the 
functions $\smash{{\cal F}^{(n)}_{\rm ion}}$ of Section \ref{sec:radiative_bias}. These will still be dimensionless functions of $k^2\lambda^2_{\rm eff}$ 
(moreover, they now depend also on the ratio $\widetilde{E}/E_\infty$). Their behavior with $k$ is the same as those of \eq{functions_f_n}: 
they go to $1$ at small $k$ and vanish at large $k$. A plot of the first two functions for different values of the effective m.f.p.~is shown in \fig{functions_of_k_energy_integral}.

\subsection{Fourier transform of angle-averaged intensity}
\label{app:fourier_transform_intensity}

\noindent In this appendix we compute the dimensionless perturbation to the angle-averaged intensity $\cal I$, 
in order to make contact with the formalism of \cite{Meiksin:2018wij,Sanderbeck:2018lwc} and to 
justify the results of \tabs{conclusions_table-1}{conclusions_table-2}. 
Once we have this quantity, if we assume that the response of galaxies to the incoming radiation happens on a time scale 
$\Delta\eta\ll{\cal H}^{-1}$, we can compute the contribution $\delta_g|_{\rm ion}$ to the galaxy overdensity by integrating it over energy and time 
with a single Green's function proportional to the photoionization cross section times $b_{\cal I}(\eta)\delta(\eta-\eta')$. 

The general solution of the RT equation is given in \eq{boltzmann_solution}. 
Focusing only on inhomogeneities in the density of emitters, we have 
\begin{equation}
\label{eq:fourier_transform_intensity-A}
\begin{split}
\int\dif\vers{n}\,\delta{\cal I}&=\int\dif\vers{n}\int_0^\eta\dif\eta'\,\bigg(\frac{1+z(\eta)}{1+z(\eta')}\bigg)^3e^{{-\tau}}\delta{\cal S}\big(\eta',\vec{x}+\vers{n}(\eta-\eta'),E(\eta',\eta)\big) \\
&=\int\dif\vers{n}\int_\eta^0\dif\eta'\,\bigg(\frac{1+z(\eta)}{1+z(\eta-y)}\bigg)^3e^{{-\tau}}\frac{(\varepsilon_{\rm em}a)\big(\eta-y,E(\eta-y,\eta)\big))}{4\pi}\,\times \\
&\hphantom{\int\dif\vers{n}\int_\eta^0\dif\eta'\,\bigg(\frac{1+z(\eta)}{1+z(\eta-y)}\bigg)^3e^{{-\tau}} } \!\delta\rho_{\rm em}(\eta-y,\vec{x}+\vers{n}y) \\
&={-\int_{\abs{\vec{y}}\,\leq\,\eta}\dif^3y}\,\bigg(\frac{1+z(\eta)}{1+z(\eta-\abs{\vec{y}})}\bigg)^3e^{{-\tau}}
\frac{(\varepsilon_{\rm em}a)\big(\eta-\abs{\vec{y}},E(\eta-\abs{\vec{y}},\eta)\big))}{4\pi\abs{\vec{y}}^2}\,\times \\
&\hphantom{{-\int_{\abs{\vec{y}}\,\leq\,\eta}\dif^3y}\,\bigg(\frac{1+z(\eta)}{1+z(\eta-\abs{\vec{y}})}\bigg)^3e^{{-\tau}} } \!\delta\rho_{\rm em}(\eta-\abs{\vec{y}},\vec{x}+\vec{y})\,\,,
\end{split}
\end{equation}
where we first changed from $\eta'$ to $y=\eta-\eta'$, and then to $\vec{y}=\vers{n}y$ (with $\dif y\dif\vers{n}=\dif^3y/\abs{\vec{y}}^2$). 
The optical depth in the above equation is given by 
\begin{equation}
\label{eq:fourier_transform_intensity-B}
\begin{split}
\tau &= \int_{\eta'}^\eta\dif\eta''\,(\sigma_{\rm ab}\widebar{n}_{\rm ab}a)\big(\eta'',E(\eta'',\eta)\big) \\ 
&= \abs{\vec{y}}\int_0^1\dif u\,(\sigma_{\rm ab}\widebar{n}_{\rm ab}a)\big(\eta-u\abs{\vec{y}},E(\eta-u\abs{\vec{y}},\eta)\big)\,\,, 
\end{split}
\end{equation}
where we defined $\eta''=\eta-u(\eta-\eta')=\eta-u\abs{\vec{y}}$ to go from the first to the second equality. 

\eq{fourier_transform_intensity-A} is all we need to compute $\delta_g|_{\rm ion}$ if the Green's function of galaxies is proportional to $\delta(\eta-\eta')$. 
After we integrate it over energy with an energy-dependent weight, we can see that its structure is very similar to \eq{definition_of_G_g-A-2}, 
with the obvious difference that there we have considered only a single flash of radiation. 
Writing $\delta\rho_{\rm em}=\widebar{\rho}_{\rm em}\delta_{\rm em}$ and expanding the dimensionless fluctuation $\delta_{\rm em}$ 
as in \eq{delta_em_bias_expansion-A}, we can compute $\delta_g|_{\rm ion}$ 
by taking advantage of the fact that $\delta_{\rm em}$ is the sum of two uncorrelated fields both evolving in a scale-independent way, 
\ie~by means of a growth factor (indeed, if the stochastic term has a power spectrum which is localized in real space at all times, 
its time evolution cannot depend on $k$). More precisely, we can write $\delta_{\rm em}$ in \eq{fourier_transform_intensity-A} as 
\begin{equation}
\label{eq:fourier_transform_intensity-C}
\begin{split}
\delta_{\rm em}(\eta-\abs{\vec{y}},\vec{x}+\vec{y}) &= \frac{D_1(\eta-\abs{\vec{y}})}{D_1(\eta)}b_{\rm em}(\eta-\abs{\vec{y}})\delta(\eta,\vec{x}+\vec{y}) \\
&\;\;\;\; + \frac{D_{\epsilon_{\rm em}}(\eta-\abs{\vec{y}})}{D_{\epsilon_{\rm em}}(\eta)}\epsilon_{\rm em}(\eta,\vec{x}+\vec{y})\,\,,
\end{split}
\end{equation}
where $D_1$ and $D_{\epsilon_{\rm em}}$ are the linear growth factors for $\delta$ and $\epsilon_{\rm em}$, respectively. 

From \eq{fourier_transform_intensity-C} we see that, technically, $(D_1(\eta')/D_1(\eta))\,b_{\rm em}(\eta')$ and $D_{\epsilon_{\rm em}}(\eta')/D_{\epsilon_{\rm em}}(\eta)$ 
play the role of two different Green's functions for the deterministic and stochastic contributions. 
In this sense, the case of continuous emission and instantaneous response mirrors that of an extended response and a single flash 
of radiation discussed in Section \ref{ssec:single_flash} (compare $\smash{G^{(1)}_g(\eta,\eta')\propto\delta(\eta-\eta')}$ with \eq{delta_dirac_emissivity}). 
However, it is important to emphasize that $(D_1(\eta')/D_1(\eta))\,b_{\rm em}(\eta')$ and $D_{\epsilon_{\rm em}}(\eta')/D_{\epsilon_{\rm em}}(\eta)$ 
are very different from the Green's functions that we have considered in Section \ref{ssec:single_flash}, 
since they have nothing to do with the response of galaxies to ionizing radiation but only encode how $\delta_{\rm em}$ evolves in time. 

These calculations also show that it is still possible to arrive at a resummation of the higher-derivative terms from RT effects 
even if we include the inhomogeneities in the optical depth. More precisely, expanding $\exp({-\tau})$ at linear order in spatial fluctuations, 
similarly to what we did in \eq{sssec_generalization_tau}, and using the fact that the deterministic and stochastic parts of $\delta_{\rm ab}$ 
evolve in a scale-independent way (see \eg~\eq{fourier_transform_intensity-C} above), 
we end up with an additional ``negative'' source whose emissivity is not localized around a single time. Thus, we fall in the same situation above. 
This tells us that the only obstacle to reliably predicting the scale dependence of the bias due to RT effects is having 
a galaxy response with support also for frequencies comparable to $\cal H$ (see also \tab{conclusions_table-2}).

\subsection{Momentum density of the radiation field}
\label{app:momentum}

\noindent In this appendix we sketch how to compute the momentum density of the ionizing radiation field 
(assuming for simplicity an homogeneous optical depth: adding inhomogeneities in $\tau$ will not change these results, 
as we see for example from the first columns of \tabs{conclusions_table-1}{conclusions_table-2} and the discussion at the end 
of the previous appendix). The energy density and momentum density for the observer $U^\mu$ described in Section \ref{ssec:emission_absorption} are\footnote{Technically, 
we should consider the fact that the fluid trajectory $\vec{x}_{\rm fl}$ is different from $\vec{x}$: 
however, these corrections come in at higher orders in perturbations so we neglect them (see also the discussion in Section \ref{ssec:q_tracers}).} 
\begin{subequations}
\label{eq:energy_and_momentum}
\begin{align}
&\rho_{\cal I} = \int_0^{+\infty}\dif E\int\dif\vers{n}\,{\cal I}\,\,, \label{eq:energy_and_momentum-1} \\
&\vec{p}_{\cal I} = {-\int_0^{+\infty}}\dif E\int\dif\vers{n}\,\vers{n}\,{\cal I}\,\,. \label{eq:energy_and_momentum-2}
\end{align}
\end{subequations}
More precisely, paralleling \eq{fourier_transform_intensity-A}, we have
\begin{equation}
\label{eq:momentum-A}
\begin{split}
{-\int\dif\vers{n}\,\vers{n}\,{\cal I}}&=
-\int\dif\vers{n}\,\vers{n}\int_0^\eta\dif\eta'\,\bigg(\frac{1+z(\eta)}{1+z(\eta')}\bigg)^3e^{{-\tau}}{\cal S}\big(\eta',\vec{x}+\vers{n}(\eta-\eta'),E(\eta',\eta)\big) \\
&=-\int\dif\vers{n}\,\vers{n}\int_\eta^0\dif\eta'\,\bigg(\frac{1+z(\eta)}{1+z(\eta-y)}\bigg)^3e^{{-\tau}}\frac{(\varepsilon_{\rm em}a)\big(\eta-y,E(\eta-y,\eta)\big))}{4\pi}\,\times \\
&\hphantom{\int\dif\vers{n}\,\vers{n}\int_\eta^0\dif\eta'\,\bigg(\frac{1+z(\eta)}{1+z(\eta-y)}\bigg)^3e^{{-\tau}} } \!\rho_{\rm em}(\eta-y,\vec{x}+\vers{n}y) \\
&={\int_{\abs{\vec{y}}\,\leq\,\eta}\dif^3y}\,\bigg(\frac{1+z(\eta)}{1+z(\eta-\abs{\vec{y}})}\bigg)^3e^{{-\tau}}
\frac{(\varepsilon_{\rm em}a)\big(\eta-\abs{\vec{y}},E(\eta-\abs{\vec{y}},\eta)\big))}{4\pi\abs{\vec{y}}^3}\,\times \\
&\hphantom{{-\int_{\abs{\vec{y}}\,\leq\,\eta}\dif^3y}\,\bigg(\frac{1+z(\eta)}{1+z(\eta-\abs{\vec{y}})}\bigg)^3e^{{-\tau}} } \!\rho_{\rm em}(\eta-\abs{\vec{y}},\vec{x}+\vec{y})\,\vec{y}\,\,.
\end{split}
\end{equation}
It is straightforward to see that the same conclusions of Appendix \ref{app:fourier_transform_intensity} apply, i.e.~it is 
possible to obtain a resummation of all the higher-derivative contributions to the momentum density $\vec{p}_{\cal I} = \delta\vec{p}_{\cal I}$. 
However, for the discussion in Section \ref{sec:conclusions} it is sufficient to stop at leading order in derivatives. Expanding 
the density of emitters in spatial derivatives inside the integral, we see that 
\begin{equation}
\label{eq:momentum-B}
\delta\vec{p}_{\cal I} = \alpha\widebar{\rho}_{\cal I}\lambda_{\rm eff}\vec{\nabla}\delta + {\cal O}(\lambda_{\rm eff}^3\vec{\nabla}\nabla^2\delta)\,\,,
\end{equation}
where $\alpha$ is a time-dependent dimensionless coefficient of order $1$ and $\rho_{\cal I}$ 
is obtained from \eqsII{fourier_transform_intensity-A}{energy_and_momentum-1} by taking $\rho_{\rm em} = \widebar{\rho}_{\rm em}$.



\clearpage

\bibliographystyle{utphys}
\bibliography{refs}

\providecommand{\href}[2]{#2}\begingroup\raggedright\begin{thebibliography}{10}

\bibitem{Desjacques:2016bnm}
V.~Desjacques, D.~Jeong, and F.~Schmidt, ``{Large-Scale Galaxy Bias},''
  \href{http://dx.doi.org/10.1016/j.physrep.2017.12.002}{{\em Phys. Rept.}
  {\bfseries 733} (2018) 1--193},
\href{http://arxiv.org/abs/1611.09787}{{\ttfamily arXiv:1611.09787
  [astro-ph.CO]}}.

\bibitem{McDonald:2009dh}
P.~McDonald and A.~Roy, ``{Clustering of dark matter tracers: generalizing bias
  for the coming era of precision LSS},''
  \href{http://dx.doi.org/10.1088/1475-7516/2009/08/020}{{\em JCAP} {\bfseries
  0908} (2009) 020},
\href{http://arxiv.org/abs/0902.0991}{{\ttfamily arXiv:0902.0991
  [astro-ph.CO]}}.

\bibitem{Efstathiou:1992zz}
G.~Efstathiou, ``{Suppressing the formation of dwarf galaxies via
  photoionization},''
{\em Mon. Not. Roy. Astron. Soc.} {\bfseries 256} (1992) 43P--47P.

\bibitem{Barkana:1999apa}
R.~Barkana and A.~Loeb, ``{The photoevaporation of dwarf galaxies during
  reionization},'' \href{http://dx.doi.org/10.1086/307724}{{\em Astrophys. J.}
  {\bfseries 523} (1999) 54},
\href{http://arxiv.org/abs/astro-ph/9901114}{{\ttfamily arXiv:astro-ph/9901114
  [astro-ph]}}.

\bibitem{Schmidt:2017lqe}
F.~Schmidt and F.~Beutler, ``{Imprints of Reionization in Galaxy Clustering},''
  \href{http://dx.doi.org/10.1103/PhysRevD.96.083533}{{\em Phys. Rev.}
  {\bfseries D96} no.~8, (2017) 083533},
\href{http://arxiv.org/abs/1705.07843}{{\ttfamily arXiv:1705.07843
  [astro-ph.CO]}}.

\bibitem{Worseck:2014fya}
G.~Worseck, J.~X. Prochaska, J.~M. O'Meara, G.~D. Becker, S.~Ellison, S.~Lopez,
  A.~Meiksin, B.~M{\' e}nard, M.~T. Murphy, and M.~Fumagalli, ``{The Giant
  Gemini GMOS survey of $z_{\rm em} > 4.4$ quasars – I. Measuring the mean
  free path across cosmic time},''
  \href{http://dx.doi.org/10.1093/mnras/stu1827}{{\em Mon. Not. Roy. Astron.
  Soc.} {\bfseries 445} no.~2, (2014) 1745--1760},
\href{http://arxiv.org/abs/1402.4154}{{\ttfamily arXiv:1402.4154
  [astro-ph.CO]}}.

\bibitem{Becker:2014oga}
G.~D. Becker, J.~S. Bolton, P.~Madau, M.~Pettini, E.~V. Ryan-Weber, and B.~P.
  Venemans, ``{Evidence of patchy hydrogen reionization from an extreme
  Ly$\alpha$ trough below redshift six},''
  \href{http://dx.doi.org/10.1093/mnras/stu2646}{{\em Mon. Not. Roy. Astron.
  Soc.} {\bfseries 447} (2015) 3402},
\href{http://arxiv.org/abs/1407.4850}{{\ttfamily arXiv:1407.4850
  [astro-ph.CO]}}.

\bibitem{Gleyzes:2016tdh}
J.~Gleyzes, R.~de~Putter, D.~Green, and O.~Dor{\'e}, ``{Biasing and the search
  for primordial non-Gaussianity beyond the local type},''
  \href{http://dx.doi.org/10.1088/1475-7516/2017/04/002}{{\em JCAP} {\bfseries
  1704} no.~04, (2017) 002},
\href{http://arxiv.org/abs/1612.06366}{{\ttfamily arXiv:1612.06366
  [astro-ph.CO]}}.

\bibitem{Pritchard:2006ng}
J.~R. Pritchard, S.~R. Furlanetto, and M.~Kamionkowski, ``{Galaxy surveys,
  inhomogeneous reionization, and dark energy},''
  \href{http://dx.doi.org/10.1111/j.1365-2966.2006.11131.x}{{\em Mon. Not. Roy.
  Astron. Soc.} {\bfseries 374} (2007) 159--167},
\href{http://arxiv.org/abs/astro-ph/0604358}{{\ttfamily arXiv:astro-ph/0604358
  [astro-ph]}}.

\bibitem{Coles:2007be}
P.~Coles and P.~Erdogdu, ``{Scale-dependent Galaxy Bias},''
  \href{http://dx.doi.org/10.1088/1475-7516/2007/10/007}{{\em JCAP} {\bfseries
  0710} (2007) 007},
\href{http://arxiv.org/abs/0706.0412}{{\ttfamily arXiv:0706.0412 [astro-ph]}}.

\bibitem{Pontzen:2014ena}
A.~Pontzen, ``{Scale-dependent bias in the baryonic-acoustic-oscillation-scale
  intergalactic neutral hydrogen},''
  \href{http://dx.doi.org/10.1103/PhysRevD.89.083010}{{\em Phys. Rev.}
  {\bfseries D89} no.~8, (2014) 083010},
\href{http://arxiv.org/abs/1402.0506}{{\ttfamily arXiv:1402.0506
  [astro-ph.CO]}}.

\bibitem{Gontcho:2014nsa}
S.~Gontcho A~Gontcho, J.~Miralda-Escud{\' e}, and N.~G. Busca, ``{On the effect
  of the ionizing background on the Ly$\alpha$ forest autocorrelation
  function},'' \href{http://dx.doi.org/10.1093/mnras/stu860}{{\em Mon. Not.
  Roy. Astron. Soc.} {\bfseries 442} no.~1, (2014) 187--195},
\href{http://arxiv.org/abs/1404.7425}{{\ttfamily arXiv:1404.7425
  [astro-ph.CO]}}.

\bibitem{Beutler:2016ixs}
{\bfseries BOSS} Collaboration, F.~Beutler {\em et~al.}, ``{The clustering of
  galaxies in the completed SDSS-III Baryon Oscillation Spectroscopic Survey:
  baryon acoustic oscillations in the Fourier space},''
  \href{http://dx.doi.org/10.1093/mnras/stw2373}{{\em Mon. Not. Roy. Astron.
  Soc.} {\bfseries 464} no.~3, (2017) 3409--3430},
\href{http://arxiv.org/abs/1607.03149}{{\ttfamily arXiv:1607.03149
  [astro-ph.CO]}}.

\bibitem{Beutler:2016arn}
{\bfseries BOSS} Collaboration, F.~Beutler {\em et~al.}, ``{The clustering of
  galaxies in the completed SDSS-III Baryon Oscillation Spectroscopic Survey:
  Anisotropic galaxy clustering in Fourier-space},''
  \href{http://dx.doi.org/10.1093/mnras/stw3298}{{\em Mon. Not. Roy. Astron.
  Soc.} {\bfseries 466} no.~2, (2017) 2242--2260},
\href{http://arxiv.org/abs/1607.03150}{{\ttfamily arXiv:1607.03150
  [astro-ph.CO]}}.

\bibitem{Rybicki:2004hfl}
G.~B. Rybicki and A.~P. Lightman,
  \href{http://dx.doi.org/10.1002/9783527618170}{{\em {Radiative Processes in
  Astrophysics}}}.
\newblock Wiley-VCH, 2004.
\newblock
\url{http://www.bartol.udel.edu/~owocki/phys633/RadProc-RybLightman.pdf}.
\newblock

\bibitem{Mao:2014rja}
Y.~Mao, A.~D'Aloisio, B.~D. Wandelt, J.~Zhang, and P.~R. Shapiro, ``{Linear
  perturbation theory of reionization in position space: Cosmological radiative
  transfer along the light cone},''
  \href{http://dx.doi.org/10.1103/PhysRevD.91.083015}{{\em Phys. Rev.}
  {\bfseries D91} no.~8, (2015) 083015},
\href{http://arxiv.org/abs/1411.7022}{{\ttfamily arXiv:1411.7022
  [astro-ph.CO]}}.

\bibitem{McQuinn:2018zwa}
M.~McQuinn and A.~D'Aloisio, ``{The observable 21cm signal from reionization
  may be perturbative},''
\href{http://arxiv.org/abs/1806.08372}{{\ttfamily arXiv:1806.08372
  [astro-ph.CO]}}.

\bibitem{Zhang:2006kr}
J.~Zhang, L.~Hui, and Z.~Haiman, ``{A Linear Perturbation Theory of
  Inhomogeneous Reionization},''
  \href{http://dx.doi.org/10.1111/j.1365-2966.2006.11311.x}{{\em Mon. Not. Roy.
  Astron. Soc.} {\bfseries 375} (2007) 324--336},
\href{http://arxiv.org/abs/astro-ph/0607628}{{\ttfamily arXiv:astro-ph/0607628
  [astro-ph]}}.

\bibitem{DAloisio:2013mgn}
A.~D'Aloisio, J.~Zhang, P.~R. Shapiro, and Y.~Mao, ``{The scale-dependent
  signature of primordial non-Gaussianity in the large-scale structure of
  cosmic reionization},'' \href{http://dx.doi.org/10.1093/mnras/stt926}{{\em
  Mon. Not. Roy. Astron. Soc.} {\bfseries 433} (2013) 2900},
\href{http://arxiv.org/abs/1304.6411}{{\ttfamily arXiv:1304.6411
  [astro-ph.CO]}}.

\bibitem{Senatore:2014eva}
L.~Senatore, ``{Bias in the Effective Field Theory of Large Scale
  Structures},'' \href{http://dx.doi.org/10.1088/1475-7516/2015/11/007}{{\em
  JCAP} {\bfseries 1511} no.~11, (2015) 007},
\href{http://arxiv.org/abs/1406.7843}{{\ttfamily arXiv:1406.7843
  [astro-ph.CO]}}.

\bibitem{Mirbabayi:2014zca}
M.~Mirbabayi, F.~Schmidt, and M.~Zaldarriaga, ``{Biased Tracers and Time
  Evolution},'' \href{http://dx.doi.org/10.1088/1475-7516/2015/07/030}{{\em
  JCAP} {\bfseries 1507} no.~07, (2015) 030},
\href{http://arxiv.org/abs/1412.5169}{{\ttfamily arXiv:1412.5169
  [astro-ph.CO]}}.

\bibitem{Sakurai:1167961}
J.~J. Sakurai, {\em {Modern quantum mechanics; rev. ed.}}
\newblock Addison-Wesley, Reading, MA, 1994.
\newblock \url{https://cds.cern.ch/record/1167961}.

\bibitem{Ichiki:2011ue}
K.~Ichiki and M.~Takada, ``{The impact of massive neutrinos on the abundance of
  massive clusters},'' \href{http://dx.doi.org/10.1103/PhysRevD.85.063521}{{\em
  Phys. Rev.} {\bfseries D85} (2012) 063521},
\href{http://arxiv.org/abs/1108.4688}{{\ttfamily arXiv:1108.4688
  [astro-ph.CO]}}.

\bibitem{Castorina:2013wga}
E.~Castorina, E.~Sefusatti, R.~K. Sheth, F.~Villaescusa-Navarro, and M.~Viel,
  ``{Cosmology with massive neutrinos II: on the universality of the halo mass
  function and bias},''
  \href{http://dx.doi.org/10.1088/1475-7516/2014/02/049}{{\em JCAP} {\bfseries
  1402} (2014) 049},
\href{http://arxiv.org/abs/1311.1212}{{\ttfamily arXiv:1311.1212
  [astro-ph.CO]}}.

\bibitem{LoVerde:2014pxa}
M.~LoVerde, ``{Halo bias in mixed dark matter cosmologies},''
  \href{http://dx.doi.org/10.1103/PhysRevD.90.083530}{{\em Phys. Rev.}
  {\bfseries D90} no.~8, (2014) 083530},
\href{http://arxiv.org/abs/1405.4855}{{\ttfamily arXiv:1405.4855
  [astro-ph.CO]}}.

\bibitem{Villaescusa-Navarro:2017mfx}
F.~Villaescusa-Navarro, A.~Banerjee, N.~Dalal, E.~Castorina, R.~Scoccimarro,
  R.~Angulo, and D.~N. Spergel, ``{The imprint of neutrinos on clustering in
  redshift-space},'' \href{http://dx.doi.org/10.3847/1538-4357/aac6bf}{{\em
  Astrophys. J.} {\bfseries 861} no.~1, (2018) 53},
\href{http://arxiv.org/abs/1708.01154}{{\ttfamily arXiv:1708.01154
  [astro-ph.CO]}}.

\bibitem{Chiang:2017vuk}
C.-T. Chiang, W.~Hu, Y.~Li, and M.~Loverde, ``{Scale-dependent bias and
  bispectrum in neutrino separate universe simulations},''
  \href{http://dx.doi.org/10.1103/PhysRevD.97.123526}{{\em Phys. Rev.}
  {\bfseries D97} no.~12, (2018) 123526},
\href{http://arxiv.org/abs/1710.01310}{{\ttfamily arXiv:1710.01310
  [astro-ph.CO]}}.

\bibitem{Chiang:2018laa}
C.-T. Chiang, M.~LoVerde, and F.~Villaescusa-Navarro, ``{First detection of
  scale-dependent linear halo bias in $N$-body simulations with massive
  neutrinos},''
\href{http://arxiv.org/abs/1811.12412}{{\ttfamily arXiv:1811.12412
  [astro-ph.CO]}}.

\bibitem{Seljak:2008xr}
U.~Seljak, ``{Extracting primordial non-gaussianity without cosmic variance},''
  \href{http://dx.doi.org/10.1103/PhysRevLett.102.021302}{{\em Phys. Rev.
  Lett.} {\bfseries 102} (2009) 021302},
\href{http://arxiv.org/abs/0807.1770}{{\ttfamily arXiv:0807.1770 [astro-ph]}}.

\bibitem{Giri:2018dln}
S.~K. Giri, A.~D'Aloisio, G.~Mellema, E.~Komatsu, R.~Ghara, and S.~Majumdar,
  ``{Position-dependent power spectra of the 21-cm signal from the epoch of
  reionization},''
\href{http://arxiv.org/abs/1811.09633}{{\ttfamily arXiv:1811.09633
  [astro-ph.CO]}}.

\bibitem{Meiksin:2018wij}
A.~Meiksin and M.~McQuinn, ``{Time-dependent fluctuations in the metagalactic
  photoionization background},''
\href{http://arxiv.org/abs/1809.08645}{{\ttfamily arXiv:1809.08645
  [astro-ph.CO]}}.

\bibitem{Sanderbeck:2018lwc}
P.~U. Sanderbeck, V.~Ir{\v s}i{\v c}, M.~McQuinn, and A.~Meiksin, ``{Estimates
  for the impact of Ultraviolet Background fluctuations on galaxy clustering
  measurements},''
\href{http://arxiv.org/abs/1810.12321}{{\ttfamily arXiv:1810.12321
  [astro-ph.CO]}}.

\end{thebibliography}\endgroup

\end{document}